\documentclass{aa}  

\usepackage{graphicx}
\usepackage{amsmath}	
\usepackage{amssymb}	
\usepackage{rotating}
\usepackage{color}

\usepackage{txfonts}
\begin{document}

\title{The gamma-ray / infrared luminosity correlation of star-forming galaxies}
\subtitle{}

\author{P. Kornecki\inst{1}, L.~J. Pellizza\inst{2}, S. del Palacio\inst{1}, A.~L. Müller\inst{1,3,4}, J.~F. Albacete-Colombo\inst{5}, and G.~E. Romero\inst{1}}


\institute{Instituto Argentino de Radioastronomía (IAR, CCT La Plata, CONICET/CIC), C.C.5, (1984) Villa Elisa, Buenos Aires, Argentina \and
Instituto de Astronomía y Física del Espacio, CONICET-UBA, C.C. 67, Suc. 28, 1428 Buenos Aires, Argentina \and
Institute for Nuclear Physics (IKP), Karlsruhe Institute of Technology (KIT), Germany \and
Instituto de Tecnologías en Detección y Astropartículas (CNEA, CONICET, UNSAM), Buenos Aires, Argentina \and Universidad de Río Negro, Sede Atlántica -- CONICET, Viedma CP8500, Río Negro, Argentina}

   \date{Received; accepted}

\authorrunning{P. Kornecki et. al}
\titlerunning{$\gamma$-ray emission of SFGs}
 
  \abstract
   {Near a dozen star-forming galaxies have been detected in $\gamma$ rays by the \textit{Fermi} observatory in the last decade. A remarkable property of this sample is the quasi-linear relation between the $\gamma$-ray luminosity and the star formation rate, obtained assuming that the latter is well traced by the infra-red luminosity of the galaxies. The non-linearity of this relation has not been fully explained yet.}
   {We aim at determining the biases derived from the use of the infra-red luminosity as a proxy for the star formation rate, and shed light onto the more fundamental relation between the latter and the $\gamma$-ray luminosity. We expect to quantify and explain some trends observed in this relation.}
   {We compile from the literature a near-homogeneous set of distances, ultraviolet, optical, infra-red, and $\gamma$-ray fluxes for all known $\gamma$-ray emitting star-forming galaxies. From these data we compute the infra-red and $\gamma$-ray luminosities, and star formation rates. We determine the best-fitting relation between the latter two, and describe the trend using simple, population-oriented models for cosmic-ray transport and cooling.}
   {We find that the $\gamma$-ray luminosity--star formation rate relation obtained from infra-red luminosities is biased to shallower slopes. The actual relation is steeper than previous estimates, having a power-law index of $1.35 \pm 0.05$, in contrast to $1.23 \pm 0.06$.}
   {The unbiased $\gamma$-ray luminosity--star formation rate relation can be explained at high star formation rates by assuming that the cosmic ray cooling region is kiloparsec-sized, and pervaded by mild to fast winds. Combined with previous results about the scaling of wind velocity with star formation rate, our work provides support to advection as the dominant cosmic ray escape mechanism in low-star formation rate galaxies.}

   \keywords{Galaxies: starburst ---
                galaxies: star formation ---
                gamma rays: galaxies
               }

   \maketitle

\section{Introduction}

Among the $\gamma$-ray sources identified with extragalactic objects \citep[e.g.,][]{4FGL2020}, those associated with star-forming galaxies (SFGs) are of particular interest. Although the sample is still small, a clear correlation is observed between the $\gamma$-ray luminosity $L_\gamma$ of these sources and observational tracers of the star formation rate (SFR) of their associated galaxies 
\citep[mainly the infrared luminosity $L_\mathrm{IR}$;][]{2012ApJ...755..164A,Tang_2014,2016MNRAS.463.1068R,2016ApJ...821L..20P,Griffin_2016,2019A&A...621A..70P,ajello2020}. This correlation suggests that the high-energy emission is produced mainly by the stellar populations of SFGs, whereas any active galactic nucleus (AGN) eventually present would provide a minor contribution.

SFGs have been observed by \textit{Fermi} at GeV energies \citep{Abdo2010,Lenain2010,2012ApJ...755..164A, Abdo2010a}, and some of them by VERITAS and H.E.S.S. at TeV energies \citep{2009Natur.462..770V,2009Sci...326.1080A}. The sample comprises very different objects: starburst galaxies (SBGs), ultraluminous IR galaxies (ULIRGs), and normal spirals such as \object{M31}. Models predicting GeV--TeV emission of SFGs had been developed before their detection \citep[e.g.,][]{Volk1996, 1999ApJ...516..744B, 2003ApJ...586L..33R, 2005A&A...444..403D, 2008A&A...486..143P, 2009arXiv0912.3497D, 2010MNRAS.401..473R}. From the first observations, a consistent picture emerged \citep{2008A&A...486..143P,2010MNRAS.401..473R}, in which the GeV emission is dominated by hadronic interactions of cosmic rays (CRs) with interstellar protons. This process produces photons through neutral pion decay. As CRs are produced mainly by supernova remnants \citep{1985ApJ...290L...1J,2017ApJ...835...72B}, high-SFR galaxies have higher CR energy densities, and therefore are more luminous in $\gamma$ rays \citep{2012JPhCS.355a2038P}. Alternative scenarios, in which galactic-scale super-winds accelerate CRs have also been proposed \citep{2018A&A...616A..57R}.

The fraction of the CR power that is lost to high-energy photons results from the interplay of CR cooling and escape mechanisms. Many works have been devoted to investigate their relevance in individual SFGs, with different levels of detail \citep{2012ApJ...755..164A,2013ApJ...762...29L,2014ApJ...780..137Y,2017ApJ...847L..13P,2018MNRAS.474.4073W,Sudoh2018,2019MNRAS.487..168P}. The results show that at high SFRs, galaxies behave as near-perfect calorimeters, radiating almost all the CR energy. Only at low SFRs non-radiative processes would be important, but their relative contribution is still controversial. Some authors \citep[e.g.,][]{Peretti2020} propose that advection is the main non-radiative cooling mechanism down to very low SFRs, when diffusion overcomes it. Others \citep[e.g.,][]{2017ApJ...847L..13P} claim that adiabatic cooling dominates the energy losses.

From the population standpoint, the most outstanding characteristic of the high-energy emission of SFGs is its correlation with SFR indicators. Using 3-year \textit{Fermi} data, \citet{2012ApJ...755..164A} have found a quasi-linear correlation $L_\gamma \propto L_\mathrm{IR}^{1.0-1.2}$ for eight SFGs. The correlation spans more than four orders of magnitude in $L_\mathrm{IR}$, but their sample is rather small. Consequently, several works have improved the data set \citep{Tang_2014,2016MNRAS.463.1068R,2016ApJ...821L..20P,Griffin_2016,2019A&A...621A..70P}. The most comprehensive work up to date is that of \citet{ajello2020}, who compile a sample of fourteen galaxies plus undetected
objects from 10-year {\em Fermi} data, finding $L_\gamma \propto L_\mathrm{IR}^{1.23 \pm 0.06}$. 

The investigation of the physics driving this correlation is important for several reasons. First, it allows us to understand the acceleration and evolution of CRs in galaxies \citep[e.g.,][]{2018MNRAS.474.4073W,2019MNRAS.487..168P}. Second, a complete description of the correlation would allow to accurately compute the contribution of SFGs to the extragalactic background $\gamma$-ray and neutrino fluxes \citep[e.g.,][]{Sudoh2018,Peretti2020}. Finally, the escaped CRs, as well as the high-energy photons, contribute to the energy feedback of SFGs into their surrounding medium. This contribution may have been important in the early Universe, when small SFGs were abundant and their energetic feedback influenced both the thermodynamic state of the intergalactic medium and the cosmic star formation history \citep[e.g.,][]{2015MNRAS.448.3071A}.

Some issues regarding the $L_\gamma$--SFR relation and its drivers remain, however, poorly explored yet. 
A large effort has been devoted to improve $\gamma$-ray data, whereas little attention has been paid to the SFR. As noted by \citet{2017ApJ...847L..13P} and \citet{2019ApJ...874..173Z}, the $L_\mathrm{IR}$--SFR relation relies on the assumption that most of the UV light emitted by massive stars is absorbed and re-radiated in the IR by dust. This is not true, for example, for low-metallicity SFGs such as the Small Magellanic Cloud (\object{SMC}). A thorough examination of this issue in the whole sample of $\gamma$-ray emitting SFGs, which is one of the aims of this work, is important to assess the validity of conclusions derived from the $L_\gamma$--$L_\mathrm{IR}$ relation.

From the theoretical side, most previous works \citep{2013ApJ...762...29L,Sudoh2018,2019MNRAS.487..168P} focus on individual galaxies. They rely on multi-parameter models to fit the $\gamma$-ray spectra of SFGs. Each individual galaxy may require a different set of parameter values, and due to unavoidable correlations and degeneracies, it is difficult to extract a clear picture of which galaxy properties shape the $L_\gamma$--SFR relation from the combination of individual model results. 
A different insight is offered by population-oriented models \citep[e.g.,][]{2019ApJ...874..173Z}. These rely only on a few parameters that show scaling relations with the SFR \citep[e.g., the Kennicutt-Schmidt or K-S law,][]{Kennicutt98,2012ARA&A..50..531K} and which influence the physical mechanisms driving CR behaviour. Usually, they do not model in detail the CR particle distribution or $\gamma$-ray spectrum to obtain $L_\gamma$, computing instead the luminosity as an SFR-dependent fraction of the total CR energy. A third approach has been proposed by \citet{2017ApJ...847L..13P}, who use high-resolution magneto-hydrodynamical (MHD) simulations to describe the formation and evolution of individual galaxies from gas in dark matter haloes, including CR injection, cooling and escape. MHD simulations can trace the physical mechanisms behind $\gamma$-ray emission, but the accuracy of their description of galaxies as a whole is limited by the sub-grid physics employed, especially that describing star formation. 

In this paper, we combine the first two approaches, building a population-oriented model that treats in detail the physical mechanisms responsible for $\gamma$-ray emission. We improve upon previous models of this kind \citep[e.g.,][]{2019ApJ...874..173Z} by including a full computation of the CR distribution and the $\gamma$-ray spectrum \citep[similar to that of][]{2019MNRAS.487..168P}, while keeping the SFR scaling relations not present in individual-galaxy emission models. We analyse carefully the parameters and scaling relations needed to describe the regions of the galaxies where high-energy radiation is emitted, and treat the rest of the parameters as fixed population means, or typical values. 

This paper is organised as follows. In Sec.~\ref{data} we present SFR data taken from the literature for the full sample of SFGs observed up to date, plus the Milky Way (MW), and discuss the reliability of $L_\mathrm{IR}$ as an SFR tracer in this sample. In Sec.~\ref{model} we develop our population-oriented model for $\gamma$-ray emission of SFGs, and compare it with previous ones, paying special attention to the scaling relations that drive the $L_\gamma$--SFR correlation. In Sec.~\ref{results} we describe our results, which we discuss in Sec.~\ref{conclusions}, where we also present our conclusions.

\section{The $L_\gamma$--SFR correlation}
\label{data}

\begin{table*}
\caption{\label{t7} Distances, SFRs, IR and $\gamma$-ray fluxes and luminosities for all $\gamma$-ray emitting SFGs known.}
\centering
\begin{tabular}{lcccccc}
\hline\hline
Galaxy & $D_\mathrm{L}$ & $F_{\gamma}$ & $F_\mathrm{IR}$ & $\dot{M}_*$ & $\log{(L_\gamma)}$ & $\log{(L_\mathrm{IR} / \mathrm{L}_\odot)}$  \\
 & & [$0.1-100\,\mathrm{GeV}$] & [$8-1000\,\mu\mathrm{m}$]\tablefootmark{l} & & [$0.1-100\,\mathrm{GeV}$] & [$8-1000\,\mu\mathrm{m}$]\\
 & Mpc & $10^{-12}\, \mathrm{erg \, cm}^{-2} \,\mathrm{s}^{-1}$ & $10^{-9}\, \mathrm{erg \, cm}^{-2} \,\mathrm{s}^{-1}$ & $\mathrm{M}_{\odot}\, \mathrm{yr}^{-1}$ &$\mathrm{erg\, s^{-1}}$ &  \\
\hline
M31 & $0.77 \pm 0.04$\tablefootmark{a} & $2.29 \pm  0.70$\tablefootmark{f} & $127.2 \pm 6.4$ & $0.26 \pm 0.02$\tablefootmark{h} & $38.21 \pm 0.14$ & $9.37 \pm 0.05$\\
NGC 253  & $3.56 \pm 0.26$\tablefootmark{a} & $8.78 \pm 0.60$\tablefootmark{f} & $92.5 \pm 4.6$ & $5.03 \pm 0.76$\tablefootmark{h} & $40.12 \pm 0.07$ & $10.56 \pm 0.07$\\
SMC      & $0.060 \pm 0.003$\tablefootmark{a}  & $29.2 \pm 1.2$\tablefootmark{f} & $622 \pm 31$ & $0.027 \pm 0.003$\tablefootmark{i} & $37.10 \pm 0.05$ & $7.85 \pm 0.05$\\
M33      & $0.91 \pm 0.04$\tablefootmark{a} & $2.02 \pm 0.38$\tablefootmark{g} & $53.8 \pm 2.7$ & $0.29 \pm 0.02$\tablefootmark{h} & $38.30 \pm 0.09$ & $9.14 \pm 0.04$\\
NGC 1068          &$10.1 \pm 1.8$\tablefootmark{b}   &  $7.46 \pm 0.55$\tablefootmark{f} & $31.6 \pm 1.6$ & $22.7 \pm 8.1$\tablefootmark{h} & $40.96 \pm 0.16$ & $11.00 \pm 0.16$ \\
LMC      &$0.050 \pm 0.003$\tablefootmark{a}  & $195.1 \pm 8.5$\tablefootmark{f} & $6777 \pm 339$ & $0.20 \pm 0.03$\tablefootmark{i} & $37.77 \pm 0.06$ & $8.72 \pm 0.06$ \\
NGC 2146 & $17.2 \pm 3.2$\tablefootmark{c}& $1.83 \pm  0.36$\tablefootmark{f} & $13.71 \pm 0.69$ & $14.0 \pm 5.2$\tablefootmark{h} & $40.81 \pm 0.18$ & $11.10 \pm 0.16$\\
NGC 2403 & $3.18 \pm 0.18$\tablefootmark{a}  & $1.22 \pm 0.28$ \tablefootmark{g} & $4.73 \pm 0.24$ & $0.37 \pm 0.03$\tablefootmark{h} & $39.17 \pm 0.11$ & $9.17 \pm 0.05$ \\
M82   & $3.53 \pm 0.26$\tablefootmark{a}    & $10.36 \pm  0.52$\tablefootmark{f} & $143.6 \pm 7.2$ & $10.4 \pm 1.6$\tablefootmark{h} & $40.19 \pm 0.07$ & $10.75 \pm 0.07$\\
NGC 3424 & $25.6 \pm 1.8$\tablefootmark{d} & $1.59 \pm 0.35$ \tablefootmark{f}& $0.910 \pm 0.046$ & $1.59 \pm 0.23$\tablefootmark{j} & $41.10 \pm 0.11$ & $10.27 \pm 0.07$\\
Arp 299 & $46.8 \pm 3.3$\tablefootmark{d}  & $1.10 \pm 0.33$\tablefootmark{g} & $10.50 \pm 0.52$ & $97 \pm 14$\tablefootmark{k} & $41.46 \pm 0.14$ & $11.86 \pm 0.07$ \\
NGC 4945 & $3.72 \pm 0.27$\tablefootmark{a}  & $11.51 \pm  0.79$\tablefootmark{f} & $63.6 \pm 3.2$ & $1.22 \pm 0.16$\tablefootmark{i} & $40.28 \pm 0.07$ & $10.44 \pm 0.07$\\
Circinus & $4.21 \pm 0.70$\tablefootmark{e} & $7.1 \pm  1.2$\tablefootmark{f} & $29.8 \pm 1.5$ & $2.05 \pm 0.63$\tablefootmark{i} & $40.18 \pm 0.16$ & $10.22 \pm 0.15$\\
Arp 220 & $80.9 \pm 5.7$\tablefootmark{d}  & $2.91 \pm  0.48$\tablefootmark{f} & $7.80 \pm 0.39$ & $214 \pm 32$\tablefootmark{k} & $42.36 \pm 0.09$ & $12.20 \pm 0.07$\\
Milky Way  & * & * & *& $ 1.90\pm 0.04$\tablefootmark{m} & $38.91 \pm 0.13$\tablefootmark{n} & $10.15 \pm 0.21$\tablefootmark{n}\\
\hline\hline
\end{tabular}
\tablefoot{
\tablefoottext{a}{\citet{2016AJ....152...50T}.}
\tablefoottext{b}{\citet{2011A&A...532A.104N}.}
\tablefoottext{c}{\citet{1988cng..book.....T}.}
\tablefoottext{d}{Derived from NED.}
\tablefoottext{e}{\citet{2009AJ....138..323T}.}
\tablefoottext{f}{\citet{4FGL2020}.}
\tablefoottext{g}{Derived from \citet{ajello2020}.}
\tablefoottext{h}{Computed from FUV \citep{2007ApJS..173..185G} + IRAS $25\, \mu\mathrm{m}$ \citep{2003AJ....126.1607S} fluxes.}
\tablefoottext{i}{Computed from FUV \citep{2012A&A...544A.101C} + IRAS $25\, \mu\mathrm{m}$ \citep{2003AJ....126.1607S} fluxes.}
\tablefoottext{j}{Computed from H$\alpha$ \citep{2008ApJS..178..247K} + IRAS $25\, \mu\mathrm{m}$  \citep{2003AJ....126.1607S} fluxes.}
\tablefoottext{k}{Computed from $F_\mathrm{IR}$.}
\tablefoottext{l}{Computed from IRAS 12, 25, 60, and $100\,\mu\mathrm{m}$ fluxes of \citet[for Circinus]{2008ApJS..178..280B} and \citet[for the remaining galaxies]{2003AJ....126.1607S}.}
\tablefoottext{m}{\citet{2011AJ....142..197C}.}
\tablefoottext{n}{\citet{strong2010}.}
}
\end{table*}

This section aims to revisit the $L_\gamma$--$L_\mathrm{IR}$ relation paying special attention to its lower end.
We construct the largest possible subsample of $\gamma$-ray emitting SFGs with near-homogeneous $\gamma$-ray, IR, and SFR data, and derive the more fundamental $L_\gamma$--SFR relation. The latter allows us to quantify the biases introduced by the use of $L_\mathrm{IR}$ as a proxy for the SFR in the whole range.

Previous works have compiled samples of $L_\mathrm{IR}$ and $L_\gamma$ published in the last 25 years. Improvements in extragalactic distance measurements in this period have produced changes in the accepted distances to many nearby galaxies, sometimes by large amounts (see, e.g., the compilation of measurements in the NASA/IPAC Extragalactic Database, NED\footnote{\tt https://ned.ipac.caltech.edu}). In some previous works, updated distance values are used to compute $L_\gamma$, whereas data on $L_\mathrm{IR}$ are taken from literature sources that use outdated distances. We enforce the self consistency of our sample by taking only fluxes from $\gamma$-ray and IR catalogues, and computing luminosities using a single, near-homogeneous set of distances. Also SFRs are computed from fluxes \citep[e.g.][]{2012ARA&A..50..531K}, therefore to preserve the self consistency of our sample we always use the same set of distances described above. The self-consistent use of distances is crucial to obtain robust results. Table~\ref{t7} compiles the estimates of $L_\gamma$, $L_\mathrm{IR}$, and $\dot{M}_*$ (the SFR) for the fourteen SFGs detected in $\gamma$-rays so far, plus the MW. In every case, we take care of computing reliable values of the uncertainties for all quantities, in order to perform meaningful statistical analyses on our sample.

\subsection{Distances}
\label{distances}

We adopt the best luminosity-distance value $D_\mathrm{L}$ available for each galaxy to construct our set of distances. We take distances from the Cosmicflows-3 catalogue of \citet{2016AJ....152...50T} when possible (nine objects). The estimates of these authors rely on redshift-independent methods (Tully-Fisher, Cepheids, red giant branch tip, etc.) therefore, to keep the homogeneity of our sample, for the remaining galaxies we searched in NED for measurements performed with similar methods, and compatible with the current cosmological model. This search failed in only three cases (\object{NGC 3424}, \object{Arp 220} and \object{Arp 299}). For these we take distances computed by NED using redshifts and Hubble flow modelling. As these are distant systems ($> 25\,\mathrm{Mpc}$), we expect redshift-dependent methods to provide reliable data.

The values obtained from the aforementioned sources are proper (metric) distances. For all but the three farthest galaxies in our sample, proper and luminosity distances coincide to better than 0.5\%, therefore we take them to be identical. In the remaining cases, we compute luminosity distances from proper ones using the 9-yr WMAP cosmological model \citep{2013ApJS..208...19H}.

\subsection{Luminosities and SFRs}

In previous works, the total IR luminosity $L_\mathrm{IR}$ in the $8-1000\,\mu\mathrm{m}$ band is usually used as a proxy for the SFR. $L_\mathrm{IR}$ can be computed from IRAS 12, 25, 60, and $100\, \mu\mathrm{m}$ fluxes using the formulae in \citet{Sanders1996}. We take IRAS fluxes from the catalogues of \citet[for \object{Circinus}]{2008ApJS..178..280B} and \citet[for all remaining galaxies except the MW]{2003AJ....126.1607S}, and derive both $F_\mathrm{IR}$ and $L_\mathrm{IR}$, the latter using the luminosity distance described in Sec.~\ref{distances}. For the MW, we adopted the IR luminosity from \citet{strong2010}.

For all the galaxies in our sample, we compute their $\gamma$-ray luminosities using the observed flux in the $0.1-100\,\mathrm{GeV}$ band, and the luminosity distance to each one. For eleven galaxies we use the most recent data provided by the \textit{Fermi} source catalogue \citep[4FGL,][]{4FGL2020}. The $\gamma -$ray flux for the Large Magellanic Cloud (\object{LMC}) in the 4FGL is the result of the sum of the fluxes of four extended sources \citep[\object{4FGL J0500.9--6945e}, \object{4FGL J0519.9--6845e}, \object{4FGL J0530.0--6900e}, and \object{4FGL J0531.8--6639e};][]{Ackermann2016}. Fluxes in the $0.1-100\,\mathrm{GeV}$ band are not reported in 4FGL for \object{Arp 299}, \object{M33}, and \object{NGC 2403}, therefore we compute them from the $0.1-800\,\mathrm{GeV}$ energy fluxes and the best-fitting spectral energy distributions (SEDs) provided by \citet{ajello2020}. We also include the modelled $\gamma$-ray luminosity for the MW \citep{strong2010,2012ApJ...755..164A}. It is important to stress that MW data were added only to perform qualitative comparisons, because the MW is an important landmark for any study of SFGs. MW data depend on models of CR propagation, $\gamma$-ray emission, and the IR interstellar radiation field. Their inclusion in the quantitative analysis would destroy the homogeneity of our data set, introducing biases that we cannot quantify.

As we aim to test the reliability of $L_\mathrm{IR}$ as a proxy for the SFR within our sample, we avoid using the \textit{total} IR luminosity in the determination of the SFR of our galaxies whenever possible. Far ultraviolet (FUV, $\lambda \sim 150\,\mathrm{nm}$) and H$\alpha$ fluxes are good tracers of SFR in unobscured systems, and several methods have been devised to include the effects of dust in obscured objects, using multi-wavelength (FUV+IR or H$\alpha$+IR) composite tracers \citep{2012ARA&A..50..531K}. Among the latter, we choose those using monochromatic IR fluxes, to keep our estimates independent of $L_\mathrm{IR}$. We compile $GALEX$ FUV fluxes \citep{2007ApJS..173..185G,2012A&A...544A.101C} and H$\alpha$ fluxes \citep{2008ApJS..178..247K} for eight and nine galaxies, respectively. Five of them have estimates of both fluxes. We combine FUV and H$\alpha$ fluxes with $IRAS$ $25\,\mu\mathrm{m}$ fluxes from \citet{2003AJ....126.1607S} and compiled distances to compute SFRs using the formulae provided by \citet{2012ARA&A..50..531K},

\begin{eqnarray}
    \log \dot{M}_* [\mathrm{M}_\odot\,\mathrm{yr}^{-1}] = \log (L_\mathrm{FUV} + 3.89 L_{25\,\mu\mathrm{m}}) [\mathrm{erg\,s}^{-1}] - 43.35,\\
    \log \dot{M}_* [\mathrm{M}_\odot\,\mathrm{yr}^{-1}] = \log (L_{\mathrm{H}\alpha} + 0.02 L_{25\,\mu\mathrm{m}}) [\mathrm{erg\,s}^{-1}] - 41.27.
\end{eqnarray}

\noindent
In these equations, $L_{\mathrm{FUV}}$ and $L_{25\,\mu\mathrm{m}}$ are the monochromatic ($\nu L_\nu$) luminosities at $150\,\mathrm{nm}$ and $25\,\mu\mathrm{m}$, respectively, and $L_{\mathrm{H}\alpha}$ is the total luminosity in the H$\alpha$ line.
For the five galaxies with both FUV and H$\alpha$ data, the two SFRs agree within measurement errors. We took the former values in the subsequent analysis. 

For two galaxies (\object{Arp~220} and \object{Arp~299}) we could not find any H$\alpha$ or FUV measurement. These are IR-luminous systems with a large obscuration, for which $L_\mathrm{IR}$ is expected to be the best SFR tracer. Therefore we compute their SFRs as

\begin{equation}
\label{sfrfromir}
    \dot{M}_*[\mathrm{M}_\odot\,\mathrm{yr}^{-1}] = 1.7 \times 10^{-10} \epsilon \, L_\mathrm{IR} [\mathrm{L}_\odot], 
\end{equation}

\noindent
where $\epsilon$ depends on the initial mass function \citep[IMF,][]{1998ApJ...498..541K}. As the FUV and optical tracers taken from \citet{2012ARA&A..50..531K} are consistent with a \citet{2003ApJ...586L.133C} IMF, we adopt $\epsilon = 0.79$ \citep{2010MNRAS.407.1403C}. Finally, following \citet{2012ARA&A..50..531K}, we take the SFR value of \citet{2011AJ....142..197C} for the MW.

To assess the reliability of our SFR values, we compare them with (non-IR-based) measurements in the literature \citep{1997A&A...328..471I,2004A&A...414...69W,2004AJ....127.1531H,2009AJ....138.1243H,2012MNRAS.425.1934F,2013ApJ...768...90L,2019ApJ...872...16D}. We find that, once rescaled to our distances, published values agree well with ours within uncertainties. The only exception is \object{NGC~3424}, for which we find a $3.1\, \sigma$ difference between our SFR and that published by \citet{2013ApJ...768...90L}, where $\sigma$ is the combined error of both measurements. Our value is higher by $\sim 0.9\,\mathrm{M}_\odot\,\mathrm{yr}^{-1}$, similar to the offset they find between their SED-based and FUV-based SFRs. These authors explain it as an effect of the different SFR time scales probed by their method and FUV-based techniques. We conclude that our sample is well suited for the task of investigating the reliability of the $L_\gamma$--$L_\mathrm{IR}$ correlation.

 \begin{figure}
  \centering
  \includegraphics[width=0.49 \textwidth]{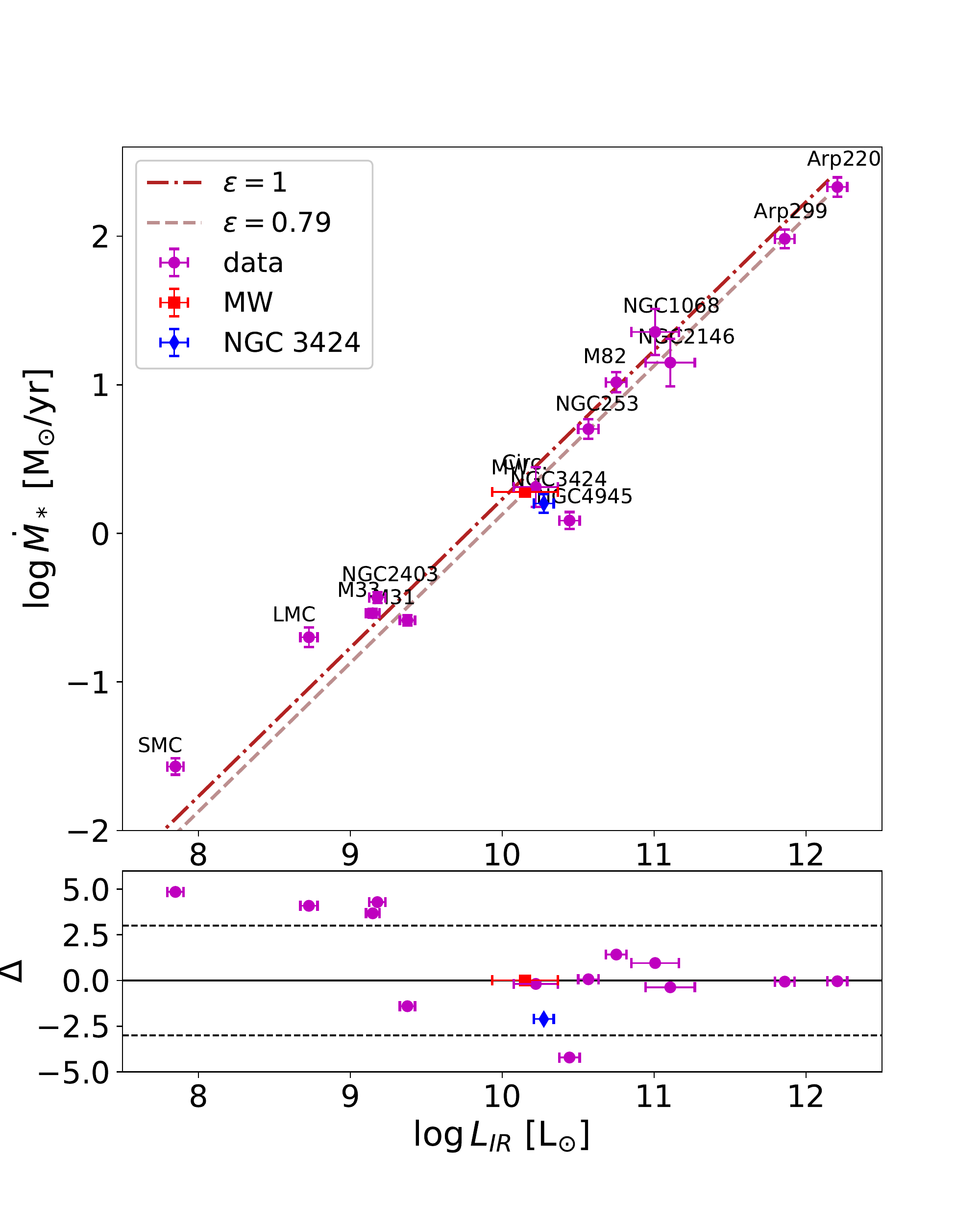}
\caption{\textit{Upper panel: SFR as a function of IR luminosity} for our sample galaxies. The red dot-dashed and grey dashed lines represent the $\dot{M}_*(L_\mathrm{IR})$ scaling relation presented in Eq.~\ref{sfrfromir} for $\epsilon = 1$ and 0.79, respectively. \textit{Lower panel:} Residuals of the $\dot{M}_*(L_\mathrm{IR})$ relation for $\epsilon = 0.79$, weighted by their standard deviation, as a function of $L_\mathrm{IR}$. Horizontal dotted lines correspond to $\pm 3$ standard deviations. $L_\mathrm{IR}$ is a good SFR tracer for $\dot{M}_* \gtrsim 1\, \mathrm{M}_{\odot} \, \mathrm{yr}^{-1} $, as expected. At the lower end, our SFRs are consistently higher than those predicted by IR luminosities.}
  \label{lgammavslir}
\end{figure}

 \begin{figure*}
  \centering
  \includegraphics[width=0.49 \textwidth]{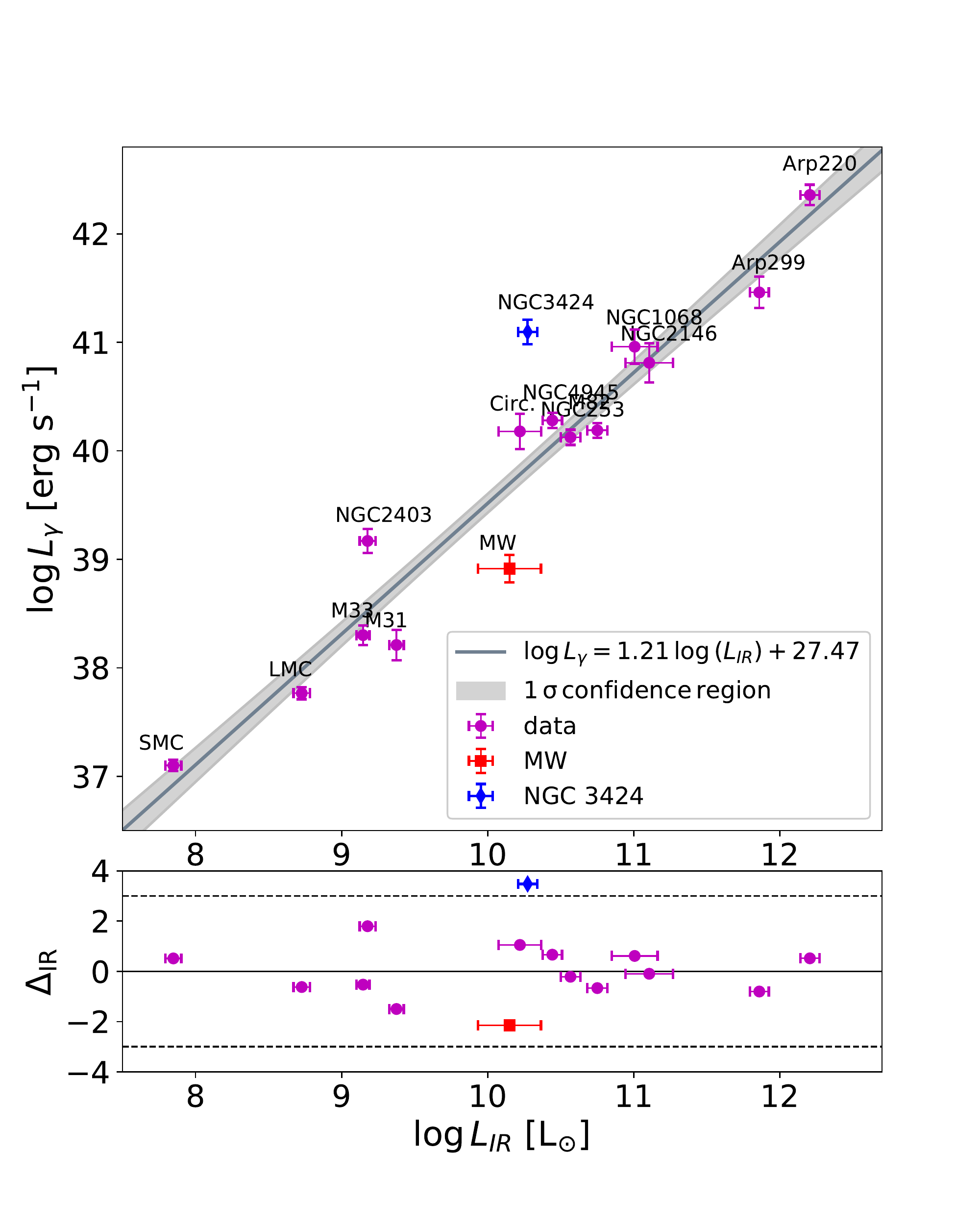}
  \includegraphics[width=0.49
  \textwidth]{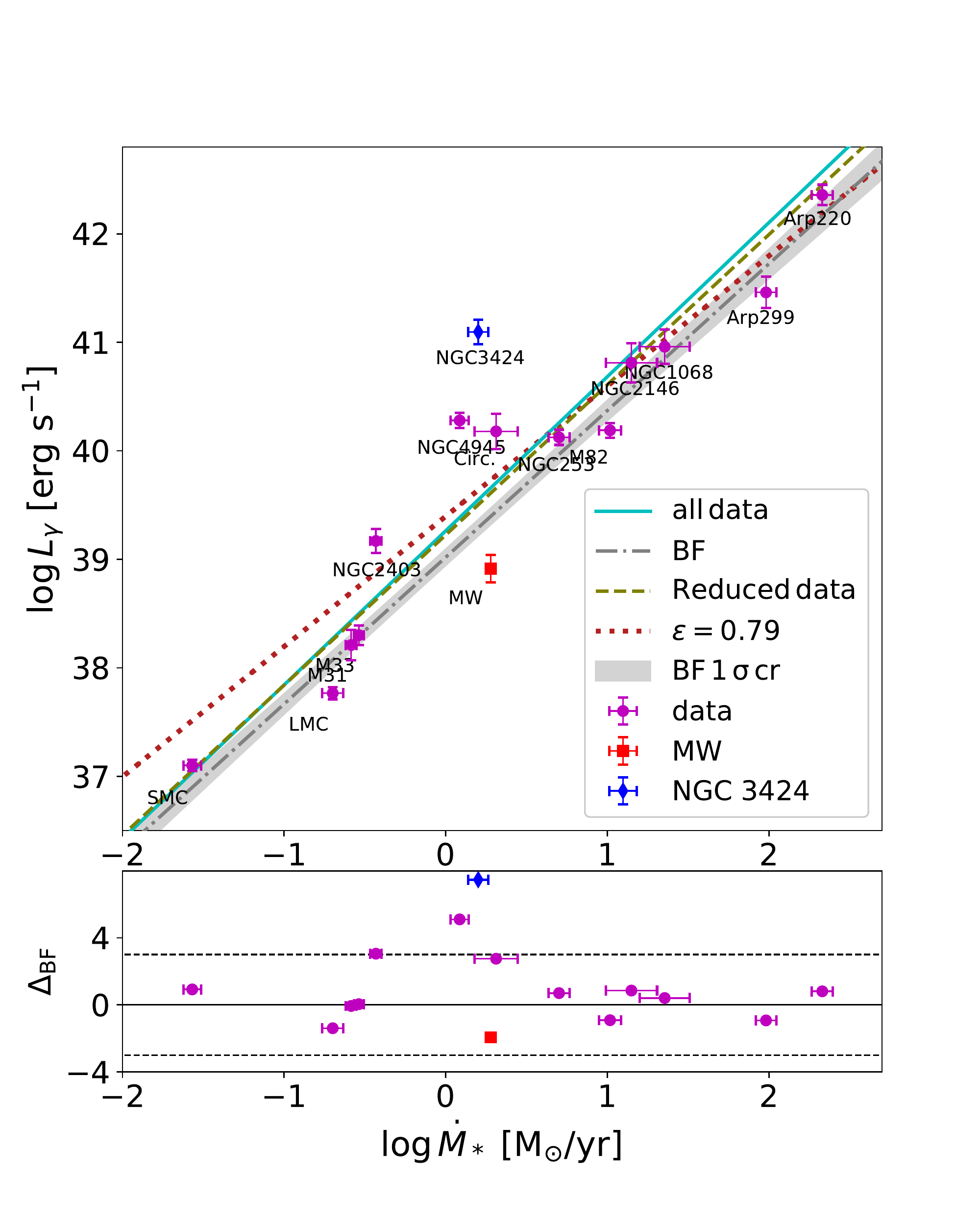}
\caption{\textit{Upper-left panel}: $\gamma$-ray luminosity of as a function of total IR luminosity for our sample. The grey solid line represents the best fit to the data (excluding the MW and \object{NGC 3424}) and the grey shaded region shows its 68$\%$ confidence level. The trend is consistent with those found by previous works. \textit{Lower-left panel}: Residuals of the fit, weighted by their standard deviation, as a function of the total IR luminosity. Horizontal dotted lines correspond to $\pm 3$ standard deviations. \textit{Upper-right panel:} $\gamma$-ray luminosity as a function of SFR. We show three fits performed on different data sets: the full data set (cyan solid line), a reduced data set (excluding the MW and \object{NGC 3424}, green dashed line), and the BF (excluding all galaxies in the range $-0.5 < \log \dot{M}_* < 0.5$, grey dot-dashed line, see text). The grey shaded region shows the 68$\%$ confidence level of the BF fit. The red dotted line represents the best fit to the $L_\gamma$--$L_\mathrm{IR}$ relation presented in the left panel, translated into the SFR--$L_\gamma$ plane by Eq.~\ref{sfrfromir}, with $\epsilon = 0.79$. The latter agrees only marginally with the data or any of the other three fits. \textit{Lower-right panel}: Residuals of the BF fit, weighted by their standard deviation, as a function of the SFR. Horizontal dotted lines correspond to $\pm 3$ standard deviations.}
  \label{lgammavsirsfr}
\end{figure*}

In the upper panel of Fig.~\ref{lgammavslir} we show the correlation we find between the IR luminosity and SFR of the galaxies of our sample, whereas in the lower panel we plot the residuals $\Delta$ of the correlation, weighted by their standard deviation (computed from both $L_\mathrm{IR}$ and $\dot{M}_*$ uncertainties). As expected, luminous IR galaxies lie on the locus given by Eq.~\ref{sfrfromir}. For comparison, we plot the corresponding relation for both $\epsilon = 0.79$ and 1 \citep[the latter corresponds to the IMF of][]{1955ApJ...121..161S}. For $\dot{M}_* \gtrsim 1\,\mathrm{M}_\odot \, \mathrm{yr}^{-1}$, most galaxies follow tightly the SFR--$L_\mathrm{IR}$ correlation defined by \citet{2012ARA&A..50..531K}. 
Only \object{NGC 3424} and \object{NGC 4945} deviate from the relation, the former marginally ($\Delta = -2.1$). The latter deserves some discussion, because the strong downward deviation ($\Delta = -4.2$) implies a higher IR emission than that expected from a complete conversion of UV photons into IR radiation by dust, at the measured SFR value. Either \object{NGC 4945} has a strong non-thermal source of IR radiation, or the H$\alpha$+IR proxy used underestimates severely the SFR. The fact that this galaxy has a strong obscuration and is observed nearly edge-on \citep{2004ApJS..151..193S} favours the second explanation. However, multi-wavelength observations would be required to completely settle this issue.

For $\dot{M}_* \lesssim 1\,\mathrm{M}_\odot \, \mathrm{yr}^{-1}$, galaxies deviate upwards from the linear relation of Eq.~\ref{sfrfromir}. Their SFRs are consistently higher than those predicted by IR luminosities; $\Delta > 3.6$ in all cases but \object{M31}. We interpret this as an effect of the incomplete obscuration of their star-forming regions, their IR luminosities accounting only for a fraction of their SFRs. This result suggests that $L_\mathrm{IR}$ is not a reliable tracer for the low-SFR end of the sample of $\gamma$-ray emitting SFGs, and that the $L_\gamma$--SFR correlation deserves further exploration.

In Fig.~\ref{lgammavsirsfr} (left panels) we present the variation of $L_\gamma$ with $L_\mathrm{IR}$, constructed using the data in Table~\ref{t7}. A clear trend is seen, from which only \object{NGC 3424} deviates significantly. The extreme $\gamma$-ray luminosity of this galaxy has already been noted by \citet{ajello2020}, and may be due to the presence of an AGN \citep{Gavazzi2011}, although this hypothesis has not been confirmed yet. A power-law fit ($L_\gamma = A L_\mathrm{IR}^m$) to the remaining data results in $m = 1.21 \pm 0.07$, $\log A = 27.47 \pm 0.65$, and a dispersion of 0.34~dex, indicating a relatively tight relation. The index $m$ is consistent with that obtained by previous authors \citep{2012ApJ...755..164A,ajello2020}. A linear behaviour ($m = 1$) can be rejected at the $3\,\sigma$ level.

The right panels of Fig.~\ref{lgammavsirsfr} show the same correlation, but using our SFR instead of its IR tracer. A fit on the full data set (excluding the MW) gives an index $m = 1.43 \pm 0.15$, $\log A = 39.32 \pm 0.17$, and a large dispersion of 0.65~dex, two times that of the $L_\gamma$--$L_\mathrm{IR}$ correlation. Using a reduced data set (excluding \object{NGC 3424} and the MW) we get $m = 1.38 \pm 0.12$, $\log A = 39.22 \pm 0.13$, but with a smaller dispersion of 0.45~dex (still larger than that of the IR correlation). The large dispersion seems to be due to the data at intermediate SFRs ($-0.5 \lesssim \log \dot{M}_* \lesssim 0.5$), therefore we try a third sample (called ``best followers'', BF) excluding the five galaxies in this SFR range. A fit to this latter data set gives $m = 1.35 \pm 0.05$, $\log A = 39.02 \pm 0.07$ and a dispersion of 0.21~dex, implying a much tighter relation. In all cases, a linear behaviour of $L_\gamma$ with SFR can be rejected at least at the $2.8\,\sigma$ level. In the case of the tight correlation, the confidence level increases to almost $7\,\sigma$.

To compare the IR and SFR correlations involving $L_\gamma$, we show in the upper-left panel of Fig.~\ref{lgammavsirsfr} the best fit to the $L_\gamma$--$L_\mathrm{IR}$ relation, translated into the SFR--$L_\gamma$ plane by Eq.~\ref{sfrfromir}. Its index ($m = 1.21 \pm 0.05$) is only marginally consistent with the ``best followers'' one ($m = 1.35 \pm 0.05$) at the $2\,\sigma$ level, where $\sigma$ is the combined error of both indices. The IR correlation largely overestimates the $\gamma$-ray luminosity of low-SFR galaxies. This is consistent with our previous result that the IR luminosity underestimates the SFR. The IR relation index neither agrees with those of the fits performed on the other two samples, but with smaller significance.

To summarise, we have shown that in our sample the IR luminosity consistently underestimates the SFR at low values. This makes the $L_\gamma$--$L_\mathrm{IR}$ relation shallower than $L_\gamma$--SFR, the index of the former being lower by 0.14. In all cases, a linear relation between $L_\gamma$ and either $L_\mathrm{IR}$ or SFR is rejected with high confidence, at least $3\sigma$. We have also found a large dispersion in the latter at intermediate SFRs. In the next section, we devise models for $\gamma$-ray emission aimed at explaining the main features of the observed $L_\gamma$--SFR correlation.

\section{The emission model}
\label{model}

\subsection{CR injection and density within galaxies}

We aim to make progress in the understanding of the processes that shape the observed $L_{\gamma}$--SFR correlation \citep{2012ApJ...755..164A}, along the whole SFR range. This encompasses both quiescent SFGs, showing low star-formation activities spread on kiloparsec-scale discs, and the most active SBGs, with large SF activities concentrated in compact, sub-kiloparsec nuclei. The correlation connects two integrated properties of galaxies, and previous works \citep[e.g.,][]{2012ApJ...755..164A,2019ApJ...874..173Z} suggest that it is possible to describe it based on the global CR energy balance.
We choose then a leaky-box model \citep{Cowsik1967} to compute the CR populations in SFGs and their emission. Leaky-box are the simplest models available. They assume a homogeneous system in which CRs are injected at some rate, and of which CRs leak out in a finite time scale. These models do
not capture the effects of gradients within SFGs, and they lack a detailed description of mechanisms such as diffusion or advection. The average effects of these mechanisms are parametrised by the leaking time scales. In spite of these shortcomings, leaky-box models can still capture the essentials of
CR energy balance \citep[e.g.,][]{2019MNRAS.487..168P}. We stress that our models do not try to fit the detailed emission of individual galaxies. Instead, we try to reproduce the mean $L_\gamma$ of a typical galaxy, given its SFR and a few global properties that either correlate with SFR or are fixed (representing averages over the whole SFG population).

In our model, CRs of energy $E$ are assumed to be injected within the galaxies at a steady rate $Q(E)$, and cool and leak out of the CR cooling region in finite time scales $\tau_\mathrm{cool}$ and $\tau_\mathrm{esc}$, respectively. The CR distribution within the region of the galaxy in which injection and cooling occurs is then

\begin{equation}
 N_i(E) = Q_i(E)\, \tau_\mathrm{loss}(E),
\end{equation}

\noindent
where $i=\mathrm{e,p}$ stands for electrons or protons, and $\tau_\mathrm{loss}^{-1} = \tau_\mathrm{esc}^{-1} + \tau_\mathrm{cool}^{-1}$. For the injection term $Q(E)$ we adopt a power law with a quasi-exponential cut-off,

\begin{equation}
 Q_i(E)=Q_{0,i} E^{-\alpha_i} \mathrm{e}^{-(E/E_{\mathrm{max}, i})^{\delta_i}},
\end{equation}

\noindent
with normalization $Q_{0,i}$, maximum particle energy $E_{\mathrm{max},i}$, spectral index $\alpha_i$, and $\delta_\mathrm{p,e} = (1,2)$ \citep{Zirakashvili2007, Blasi2010}. We assume that CRs are accelerated in SNR shocks by the Fermi diffusive shock acceleration mechanism \citep{ Axford1977, Bell1978, Blandford1978}, leading to $\alpha_i = 2.2$ for both protons and electrons \citep{2013ApJ...762...29L}. According to \citet{Merten2017}, plausible hadron-to-lepton ratios for SFGs are $a = Q_{0,\mathrm{p}}/Q_{0,\mathrm{e}} > 10$. At these large ratios, it is expected that hadrons dominate the emission, rendering the exact value of $a$ irrelevant. We therefore adopt an intermediate value $a = 50$ in the following discussion. We have verified a posteriori that different values of $a > 10$ do not change our results.

The total CR proton power injected by SNRs into the interstellar medium (ISM) of the galaxies is

\begin{equation}
    L_{\rm CR} = \int_{E_\mathrm{min,p}}^{E_\mathrm{ max,p}} E \, Q_\mathrm{p}(E)\, dE
= \xi \, E_\mathrm{SN} \, \Gamma_\mathrm{SN},
\label{eq:Lcr}
\end{equation}

\noindent
where $\xi = 0.1$ is the injection efficiency, and $E_\mathrm{SN} = 10^{51}\,\mathrm{erg}$ is the typical energy released by a SN explosion. The supernova rate of the galaxy is $\Gamma_\mathrm{SN} = (83\, \mathrm{M}_\odot)^{-1}\, \dot{M}_*$, consistent with the \citet{2003ApJ...586L.133C} IMF adopted in this work \citep{2012ApJ...755..164A}. This provides the normalization value of $Q_{0,p}$, and gives the basic dependence of $\gamma$-ray emission on SFR. 

In the presence of Bohm diffusion, as usually assumed for SNR environments, maximum particle energies can be obtained by the formulae given by \citet{Gaisser2016}.
For a typical shock velocity of $v_\mathrm{sh} = 5000 \, \mathrm{km \, s}^{-1}$, and a typical SNR magnetic field of $B_\mathrm{SNR} = 200 \, \mu \mathrm{G}$, we obtain $E_\mathrm{max,p} \sim 1\,\mathrm{PeV}$ and $E_\mathrm{max,e} \sim 10\, \mathrm{TeV}$. 
We also adopt $E_\mathrm{min,p} = 1.2 \, \mathrm{GeV}$ and $E_\mathrm{min,e} = 1 \, \mathrm{MeV}$ as minimum particle energies.

\subsection{CR escape}

CRs are advected by supernova-driven galactic winds \citep[e.g.][]{Strickland_2009}, and diffuse in the ISM due to the interaction with magnetic turbulence. Both processes lead to the escape of CRs from the galaxy at a rate $\tau_\mathrm{esc}^{-1} = \tau_\mathrm{adv}^{-1} + \tau_\mathrm{diff}^{-1}$, where $\tau_\mathrm{diff}$ and $\tau_\mathrm{adv}$ are the diffusion and advection characteristic time scales, respectively.

For $\tau_\mathrm{adv}$ we adopt the time that takes the wind to leave the CR cooling region \citep{2012JPhCS.355a2038P},

\begin{equation}
      \tau_{\rm adv} \sim 9.8\,\times 10^8 \, \left(\frac{H}{\mathrm{kpc}}\right)\, \left(\frac{v_\mathrm{w}}{\mathrm{km\, s}^{-1}}\right)^{-1} \, \mathrm{yr}, 
      \label{eq:5}
\end{equation}

\noindent
where $v_\mathrm{w}$ is the galactic wind velocity, and $H$ the shortest size of the region (i.e., the disc height).

On the other hand, $\tau_\mathrm{diff} = H^2 /D(E)$, where $D(E)$ is the diffusion coefficient, for which we explore two prescriptions representing extreme conditions for magnetic turbulence. For the first one, we adopt a Kolmogorov diffusion coefficient with the normalization found for our Galaxy, $D_\mathrm{K} = 3.86 \times 10^{28} (E/ \mathrm{GeV})^{1/3}\, \mathrm{cm}^2\,\mathrm{s}^{-1}$ by \citet{Berezinsky1990}, which leads to a fast diffusion. More recent estimates agree in order of magnitude with this value \citep[see, e.g.,][and references therein]{Gabici2019}. For the other case, we adopt a Bohm diffusion coefficient $D_\mathrm{B} = E c /(3 e B)$, where $B$ is the magnetic field of the ISM, $e$ the electron charge, and $c$ the speed of light. The recipes for determining the values of $H$, $B$, and $v_\mathrm{w}$ will be discussed in Sec.~\ref{parameters}.

\subsection{CR cooling and emission}

Cooling of CRs proceeds by different mechanisms, each one contributing with a rate $\tau_i^{-1}$ to the total cooling rate as

\begin{equation}
\tau_\mathrm{cool}^{-1} = \sum_i \tau_i^{-1}.
\end{equation}

\noindent
The considered cooling mechanisms are synchrotron radiation, inverse Compton (IC) scattering, ionization, and Bremsstrahlung for electrons, and inelastic $p$-$p$ scattering and ionization for protons. Their time scales can be expressed as \citep[and references therein]{Ginzburg1964, Blumenthal1970, 2013ApJ...762...29L, Schlickeiser, Kelner2006},

\begin{equation}
      \tau_\mathrm{sync} \sim 1.3 \times 10^{10} \left(\frac{E}{\mathrm{GeV}}\right)^{-1} \left(\frac{B}{\mu \mathrm{G}}\right)^{-2} \mathrm{yr},
\end{equation}

\begin{equation}
      \tau_\mathrm{BS} \sim 3.9 \times 10^7 \left(\frac{n}{\mathrm{cm}^{-3}}\right)^{-1} \mathrm{yr},
\end{equation}

\begin{equation}
      \tau_\mathrm{ion,e} \sim 4.8\times 10^7 \left(\frac{n}{\mathrm{cm}^{-3}}\right)^{-1}\left(\frac{E}{\mathrm{GeV}}\right)\, \mathrm{yr},
\end{equation}

\begin{equation}
      \tau_\mathrm{ion,p} \sim 1.7 \times 10^8  \left(\frac{n}{\mathrm{cm}^{-3}}\right)^{-1} \left(\frac{E}{\mathrm{GeV}}\right)\, \beta \, \, \mathrm{yr},
      \label{ionizp}
\end{equation}

\begin{equation}
      \tau_\mathrm{pp} = (n\, c\, \sigma_\mathrm{pp}(E) \, \kappa)^{-1},
      \label{taupp}
\end{equation}

\noindent
where $n$ is the ISM proton density (whose value will be discussed in Sec.~\ref{parameters}), and $\beta = v/c$ with $v$ the CR proton velocity. The cross section for inelastic $p$-$p$ scattering is \citep{Kelner2006},

\begin{equation}
\sigma_\mathrm{pp}(E) = (34.3 + 1.88 L + 0.25 L^2) \times \left[1-\left(\frac{E_\mathrm{th}}{E} \right)^4 \right]^2 \,\mathrm{mb},
\end{equation}

\noindent
where $E_\mathrm{th} = 1.22 \times  10^{-3}\, \mathrm{TeV}$ is the threshold energy for $\pi^0$  meson production, $L = \ln (E / \mathrm{TeV})$ and  $\kappa = 0.5$ is the in-elasticity of the process. Ionization time scales assume a small fraction of ionised gas in the region where CRs propagate.

The main contribution to the photon field in the ISM is provided by cold dust that radiates in the IR, with a quasi-black-body spectrum \citep[e.g.][]{Draine2011}. Therefore, for the IC cooling time we use the parametrisation given by \citet{Khangulyan2014} for an isotropic,  diluted (by a factor $\Sigma$), black-body radiation field of temperature $T = 20\,\mathrm{K}$,

\begin{equation}
\tau_\mathrm{IC}(E) = \frac{\pi \hbar^3 E}{2 \Sigma r_0 m_\mathrm{e}^3 c^4 T^2} F_{\mathrm{iso}}^{-1},
\end{equation}

\noindent
where $\hbar$ is the reduced Planck constant, $r_0$ the classical electron radius, $m_\mathrm{e}$ the electron mass, and $F_{\mathrm{iso}}$ a dimensionless function of $T$ and $E$. The value of $\Sigma$ depends on the geometry of the system, and will be discussed in Sec.~\ref{parameters}.

The above formulae allow us to compute the particle distribution $N_\mathrm{p,e}(E)$ predicted by our model. From these, we compute the SED of each emission process, using standard formulae for their emissivities \citep{Blumenthal1970, Kelner2006, Khangulyan2014}. We obtain the total $\gamma$-ray luminosity in the \textit{Fermi} energy range by integrating the SED between $0.1 - 100\,\mathrm{GeV}$.

%
\subsection{Parameters and scale relations}
\label{parameters}
%

Our model has a single independent variable, the SFR, which defines the amount of CRs produced in each galaxy. This is linear in the SFR, therefore the non-linearity of the observed $L_\gamma$--SFR relation must arise from other SFR-dependent parameters. The most obvious candidate is the ISM proton density, as a large body of work has documented the existence of a correlation between the surface densities of SFR ($\Sigma_\mathrm{SFR}$) and cold gas mass ($\Sigma_\mathrm{gas}$) of galaxies \citep[the K-S law,][]{Kennicutt98,2012ARA&A..50..531K,2019ApJ...872...16D},

\begin{equation}
    \log{\Sigma_\mathrm{SFR}}\,[\mathrm{M}_\odot \,\mathrm{yr}^{-1}\,\mathrm{kpc}^{-2}] = 1.41 \log{\Sigma_\mathrm{gas}}\,[\mathrm{M}_\odot \,\mathrm{pc}^{-2}] - 3.74.
\end{equation}

\noindent
We include the K-S law in our model to compute the density of protons $n$. As the K-S law relates intensive quantities, we are forced to take into account the geometry of the CR acceleration and cooling region.
Most SFGs and SBGs show flattened morphologies, therefore we model this region as a disc of radius $R$ and thickness $2H$. The proton density is then

\begin{equation}
    n = \frac{\Sigma_\mathrm{gas}}{2 H m_\mathrm{H}} \propto \dot{M}_*^{0.71}\,R^{-1.42}\,H^{-1}.
    \label{eqdensity}
\end{equation}

\noindent
We stress that this disc models the region where CRs cool, which needs not be the whole galaxy. Indeed, \citet{2019ApJ...872...16D} have found that star-forming regions in disc galaxies, defined by H$\alpha$ emission, are confined to smaller radii (by a factor of almost two) than the galactic light. For this reason, we do not use empirical or simulated radius--SFR or radius--stellar mass relations to define $R$, as previous works do \citep[e.g.][]{2019ApJ...874..173Z}. Instead, we compute different scenarios of our model with fixed values of $R$, between $100\,\mathrm{pc}$ and $5\,\mathrm{kpc}$ (see Table~\ref{modelparameters}). In all cases we set $H/R = 0.2$, consistent with the thickness-to-diameter ratio of Sloan Digital Sky Survey galaxies \citep{2008MNRAS.388.1321P}. We stress that the value found by these authors refers to the size of the whole disc, whereas here $R$ is the radius of the region where CRs cool. We assume that the aspect ratio of the system is the same in both cases. 

The disc geometry also allows us to compute the dilution factor of the radiation involved in IC cooling, $\Sigma = L_{\rm IR}/(8 \pi R^2 \sigma_\mathrm{SB} T^4) $ with $\sigma_\mathrm{SB}$ the Stephan-Boltzmann constant. We compute $L_\mathrm{IR}$ from the SFR using Ec.~\ref{sfrfromir} although per our results of Sec.\ref{data}, at low SFRs this is an overestimation of both $L_\mathrm{IR}$ and the IC luminosity. As we will see in Sec.~\ref{results}, the contribution of IC to the $\gamma$-ray SED in the \textit{Fermi} band is negligible, therefore a more accurate computation of IC is pointless.

Galactic winds in SFGs are complex structures, in which components of different temperatures and ionization states are mixed. The mass and momentum outflow of each component is still poorly known, therefore it is not clear which of them would dominate CR drag. Theoretical estimates of the wind terminal velocity are of the order of $3000\,\mathrm{km\,s}^{-1}$, whereas observed component velocity spans several orders of magnitude, from $\sim 30-40\,\mathrm{km\,s}^{-1}$ to over $3000\,\mathrm{km\,s}^{-1}$ \citep[and references therein]{Veilleux2005}. A correlation of the form $v_\mathrm{w} \propto \dot{M}_*^{0.35}$ has been found for the neutral component \citep{Martin2005,Weiner2009}. This result would suggest to follow \citet{2019ApJ...874..173Z}, and introduce directly the aforementioned correlation in our model. However, it is not clear whether this component is the one driving CR advection. Therefore, we prefer to follow a different strategy: we probe several fixed values for $v_\mathrm{w}$ (from 40 to $4000\,\mathrm{km\,s}^{-1}$ to roughly match the observed range) and discuss the effects of the variation of this parameter in the model results.

Finally, the magnetic fields of the CR cooling region in SFGs are not well constrained by observations. \citet{2013A&A...555A..23A} measure fields from $\sim 20$ to $\sim 100\,\mu\mathrm{G}$ in \object{M82}, whereas authors modelling high-energy emission assume a range of values from those measured to up to two orders of magnitude higher \citep[e.g.,][]{2013ApJ...762...29L,2019MNRAS.487..168P}. We adopt a conservative approach, assuming a typical value $B = 200 \, \mu \mathrm{G}$ for SFGs, and varying it by an order of magnitude above and below to represent the present degree of uncertainty.

We compute two base scenarios (K0 and B0, respectively, for Kolmogorov and Bohm diffusion prescriptions) using the values shown in Table~\ref{modelparameters}. To assess the effects of free parameters, we vary them one at a time, constructing twelve additional scenarios (numbered K1 to K6 for Kolmogorov diffusion, and B1 to B6 for Bohm; Table~\ref{modelparameters} summarises the parameter values used). For each scenario we solve for the proton and electron distributions, and compute emission spectra and $L_\mathrm{IR}$ of twenty galaxies with different SFRs, logarithmically spaced in the range $0.005-200\, \mathrm{M}_\odot\, \mathrm{yr}^{-1}$.

\begin{table}
    \centering
    \begin{tabular}{lccc}
    \hline \hline
    Scenario & $B$ & $v_\mathrm{w}$ & $R$ \\
     & $\mu\mathrm{G}$ & $\mathrm{km\,s}^{-1}$ & kpc \\ \hline
    0 & 200 & 400 & 1 \\
    1 & 200 & 400 & 0.1 \\
    2 & 200 & 400 & 5 \\
    3 & 200 & 40 & 1 \\
    4 & 200 & 4000 & 1 \\
    5 & 20 & 400 & 1 \\
    6 & 2000 & 400 & 1 \\
        \hline \hline
    \end{tabular}
    \caption{Values of the free parameters of our model (magnetic field of the galaxy, wind velocity, and radius of the CR cooling region), for the different scenarios considered.}
    \label{modelparameters}
\end{table}

\section{Model results}
\label{results}

In this section we explore the ability of our model to reproduce the main trends observed in the $L_\gamma$--SFR relation. We analyse the dominant cooling and escape mechanisms along the SFR range, the resulting $\gamma$-ray SEDs and luminosities, and assess the validity of the calorimetric hypothesis usually invoked for SBGs.

\subsection{Energy losses}
\label{resultlosses}

 \begin{figure}[t]
  \centering
  \includegraphics[angle=270, width=0.49 \textwidth]{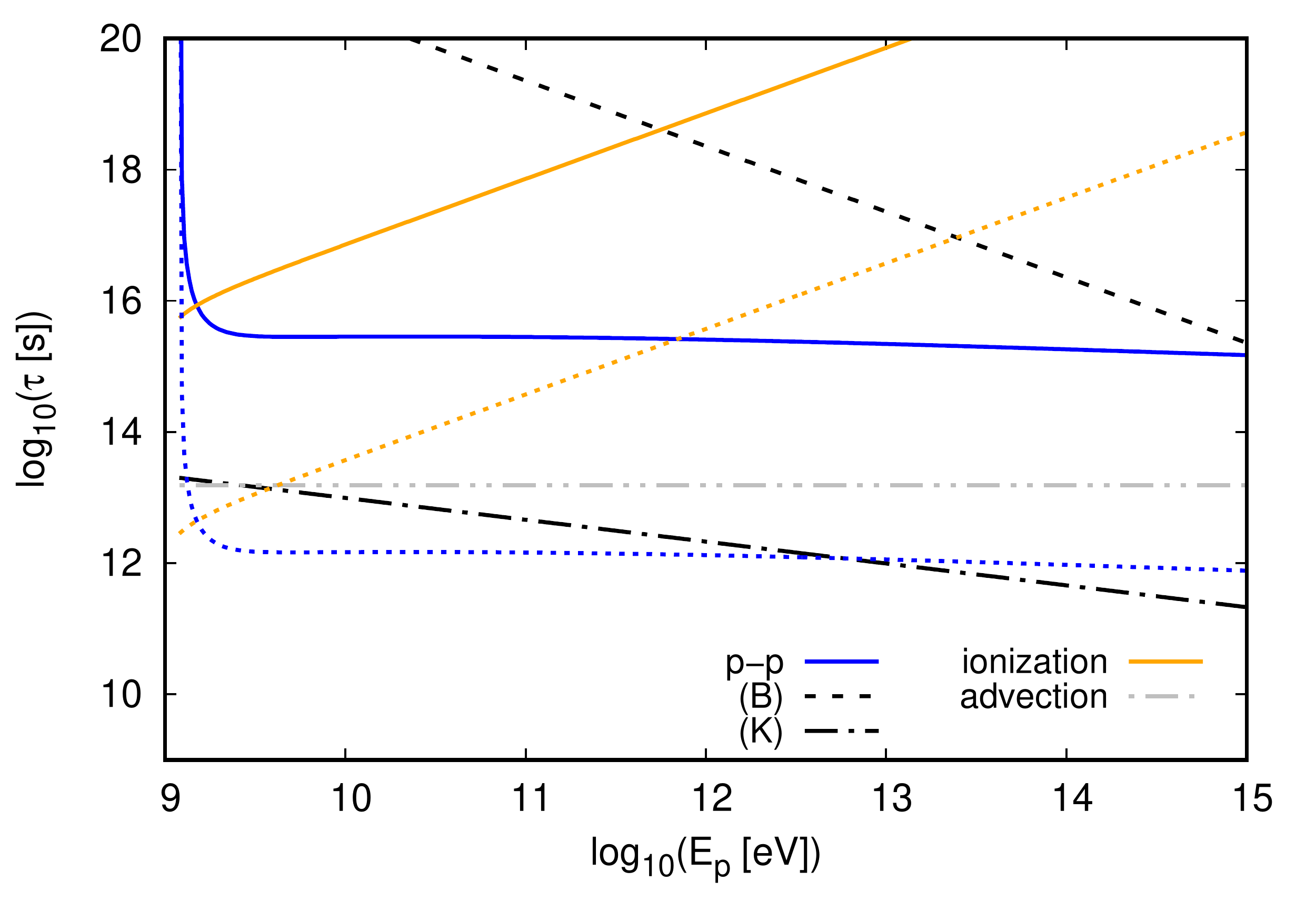}
  \caption{Cooling time scales for protons, as a function of energy, for two galaxies with $\dot{M}_* = 0.005$ (solid lines) and $200\, \mathrm{M}_{\odot} \, \mathrm{yr}^{-1}$ (dotted lines), in the base (0) scenarios. Blue and yellow lines are for $p$-$p$ scattering and ionization, respectively. Black lines represent diffusion time scales (dot-dashed and dashed for K0 and B0 scenarios, respectively), whereas the grey double-dot-dashed line is the advection time scale. At low SFRs escape dominates energy losses; as the SFR increases, more energy is radiated by $p$-$p$ mechanism, and the system approaches a calorimeter.}
  \label{protoncooling}
\end{figure}
\noindent

We show in Fig.~\ref{protoncooling} the escape and cooling times at extreme SFRs (0.005 and $200\,\mathrm{M}_\odot\,\mathrm{yr}^{-1}$), for the two base scenarios (K0 and B0). At low SFRs, escape dominates the energy losses, either through advection in the Bohm slow-diffusion scenario (B0), or through Kolmogorov diffusion (scenario K0). Protons cool via $p$-$p$, except at the lowest energies, in which ionization eventually dominates. Cooling rates are 2--4 orders of magnitude lower than escape rates (depending on energy and diffusion mode), implying that only a small fraction of the CR energy is emitted as $\gamma$ rays. As the SFR increases, $\tau_\mathrm{ion,p}$ and $\tau_\mathrm{pp}$ drop down as $\dot{M}_*^{-0.71}$, the latter dominating losses in the GeV--TeV proton energy range. More energetic protons still diffuse away in scenario K0, never reaching a calorimetric situation in the modelled region.  In scenario B0, instead, the main escape mechanism is advection, and $p$-$p$ losses dominate at high SFRs ($\dot{M}_* > 20 \,\mathrm{M}_{\odot} \, \mathrm{yr}^{-1}$) in the whole proton energy range. At SFRs of hundreds of solar masses per year, $\tau_\mathrm{adv}$ is an order of magnitude higher than $\tau_\mathrm{pp}$, therefore the calorimetric limit is approached.

 \begin{figure*}[t]
  \centering
  \includegraphics[angle=270, width=0.49 \textwidth]{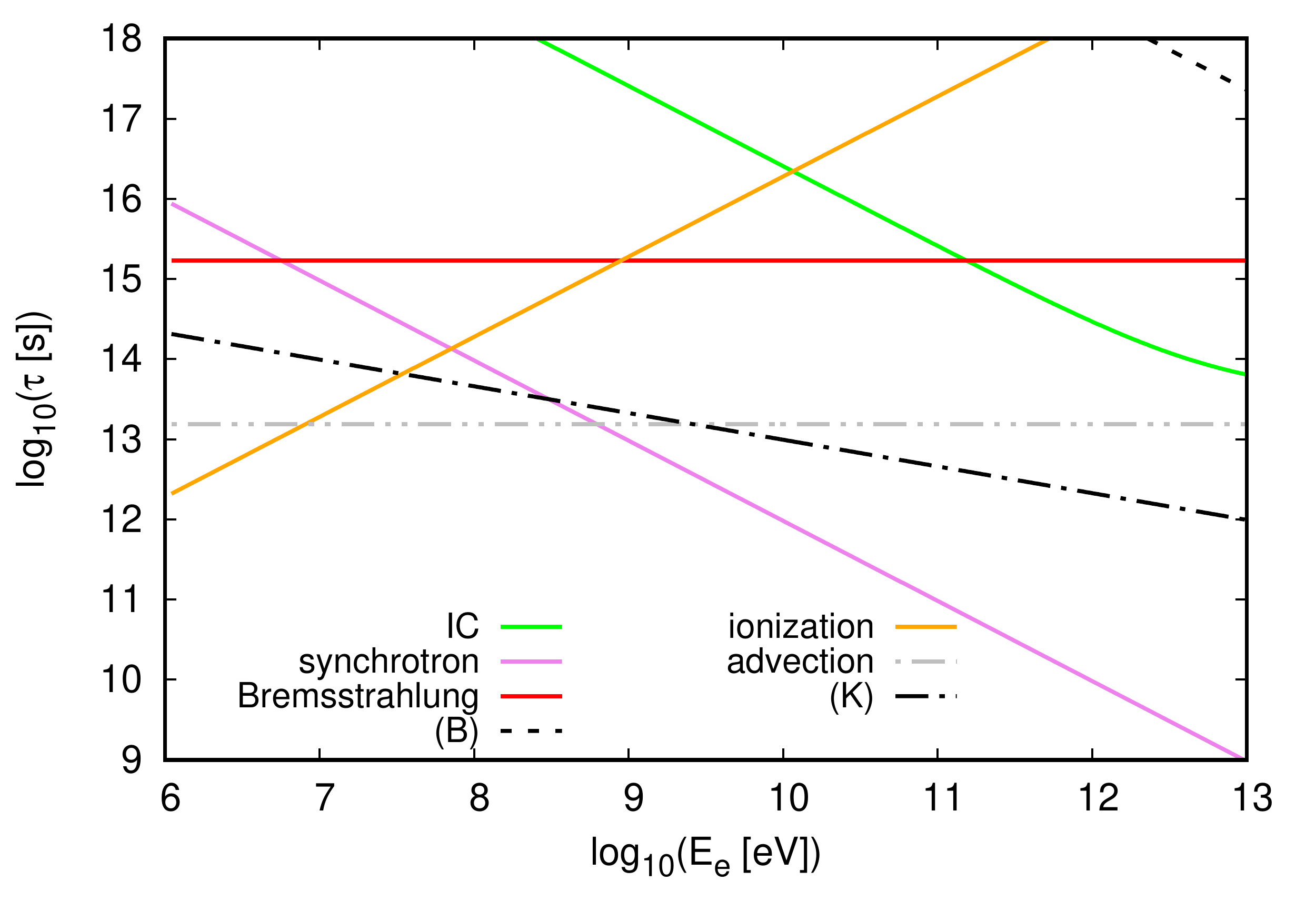}
  \includegraphics[angle=270, width=0.49 \textwidth]{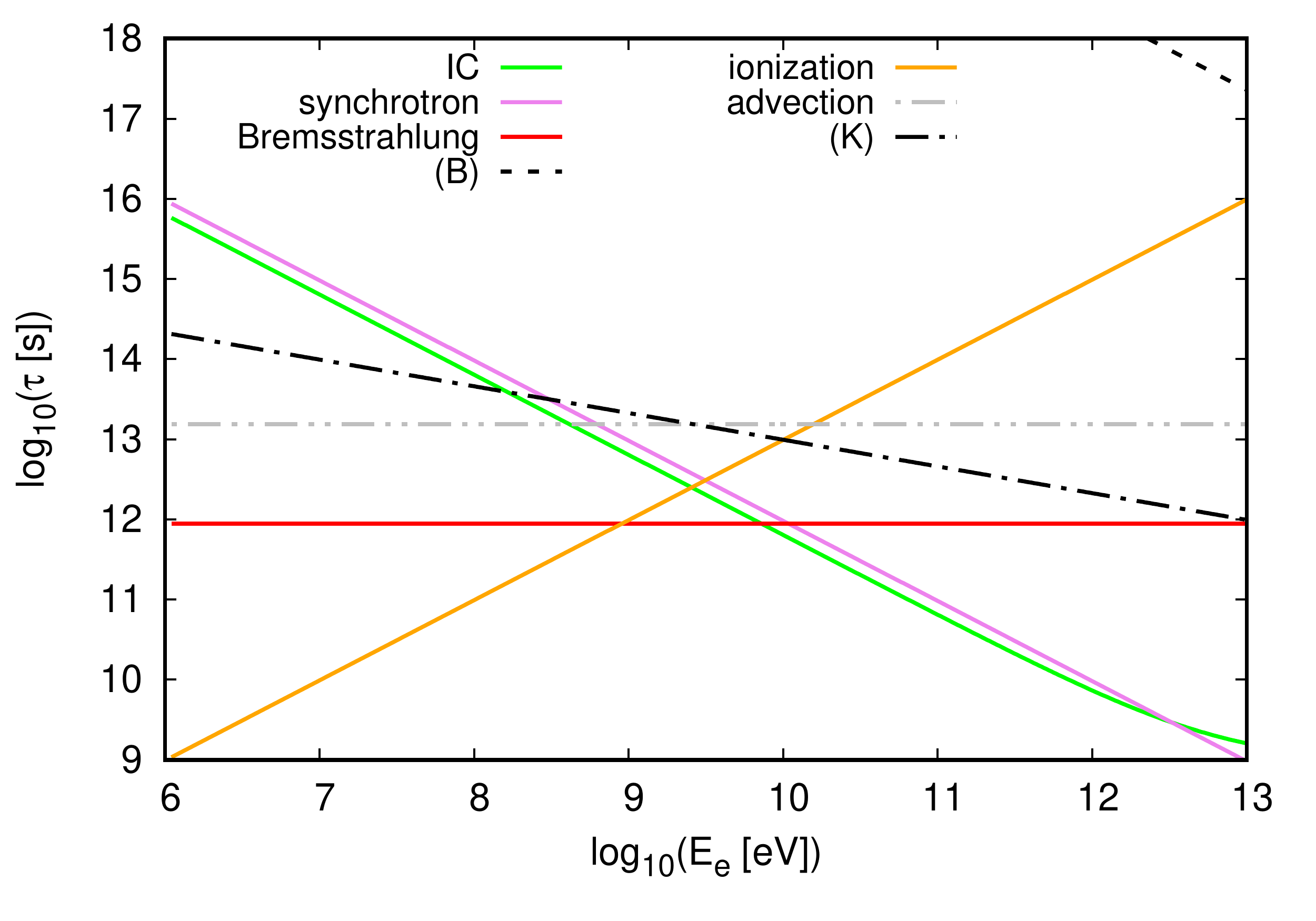} \\
  \caption{Escape and cooling time scales for electrons, as a function of energy, for two galaxies with $\dot{M}_* = 0.005$ (left panel) and $200\, \mathrm{M}_{\odot} \, \mathrm{yr}^{-1}$ (right panel), in the base (0) scenarios. Colour solid lines represent different cooling processes: ionization (yellow), Bremsstrahlung (red), IC (green), and synchrotron (pink). Black lines are for diffusion (dot-dashed and dashed for K0 and B0 scenarios, respectively), and the grey double-dot-dashed line for advection. Escape dominates losses only at low SFRs and for electron energies below $1\,\mathrm{GeV}$. At high SFRs, Bremsstrahlung and ionization cooling overtake escape losses in the same energy range. At higher energies synchrotron dominates, competing only with IC at very high SFRs.}
   \label{electroncooling}
\end{figure*}

In Fig.~\ref{electroncooling} we show the escape and cooling times for electrons, at the same extreme SFRs plotted in Fig.~\ref{protoncooling}, for the base scenarios K0 and B0. Diffusion can be neglected in both scenarios except for a small energy range around $1\,\mathrm{GeV}$ in scenario K0. Advection is the dominant energy loss mechanism for electrons at low electron energies ($\lesssim 1\, \mathrm{GeV}$), for low-SFR galaxies, whereas at higher energies electrons are cooled by synchrotron emission. As SFR increases, the time scales for Bremsstrahlung and ionization (both $\propto \dot{M}_*^{-0.7}$) decrease, and these processes become dominant below $\sim 10\, \mathrm{GeV}$. IC (with a time scale $\propto \dot{M}_*^{-1}$) gets stronger and competes with synchrotron emission at higher energies. For $\dot{M}_* \gtrsim 1 \, \mathrm{M}_{\odot} \, \mathrm{yr}^{-1}$ the system becomes a perfect electron calorimeter.

\subsection{Spectral energy distributions}
\label{resultsseds}

 \begin{figure*}[t]
  \centering
  \includegraphics[angle=270, width=0.49 \textwidth]{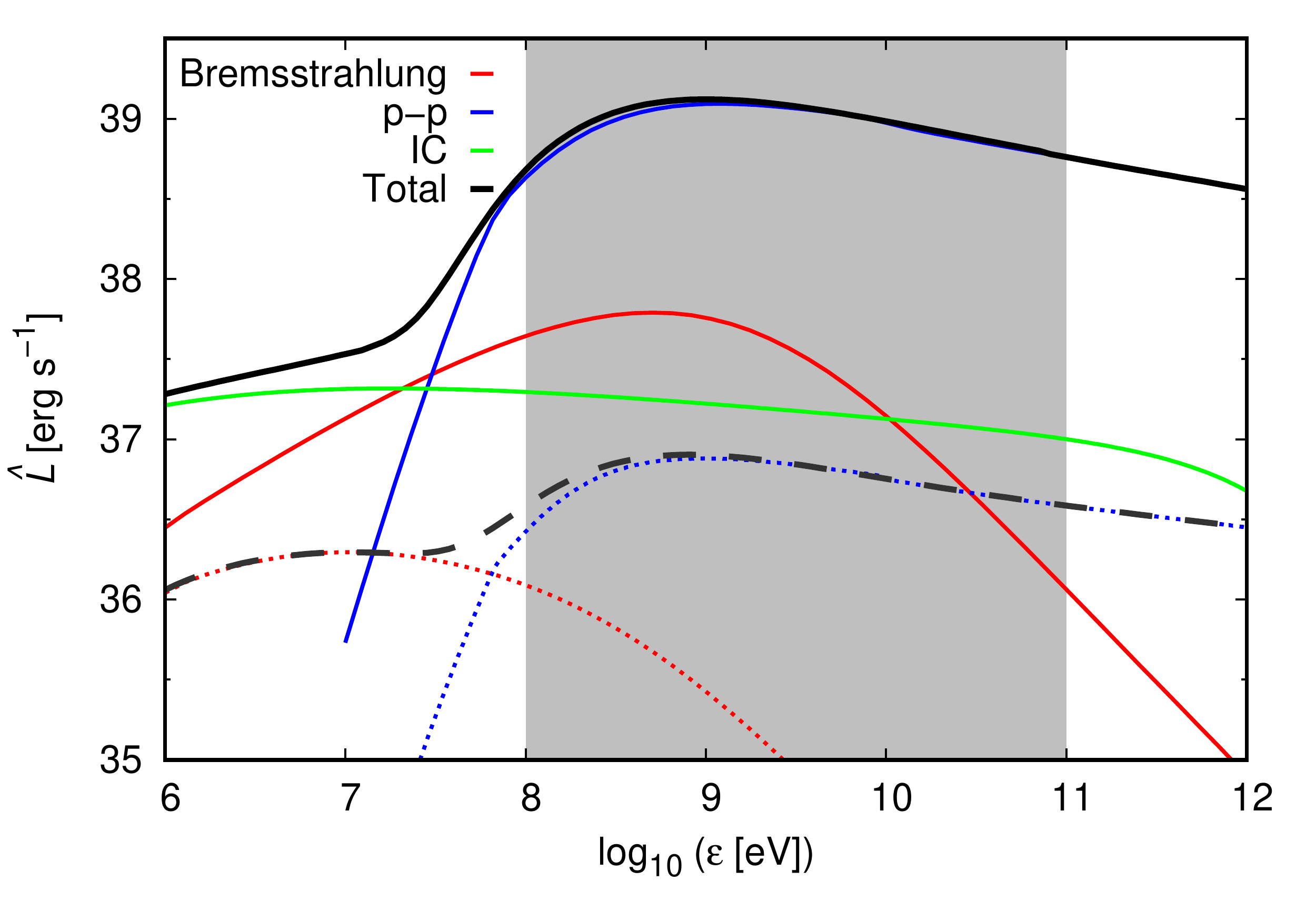}
  \includegraphics[angle=270, width=0.49 \textwidth]{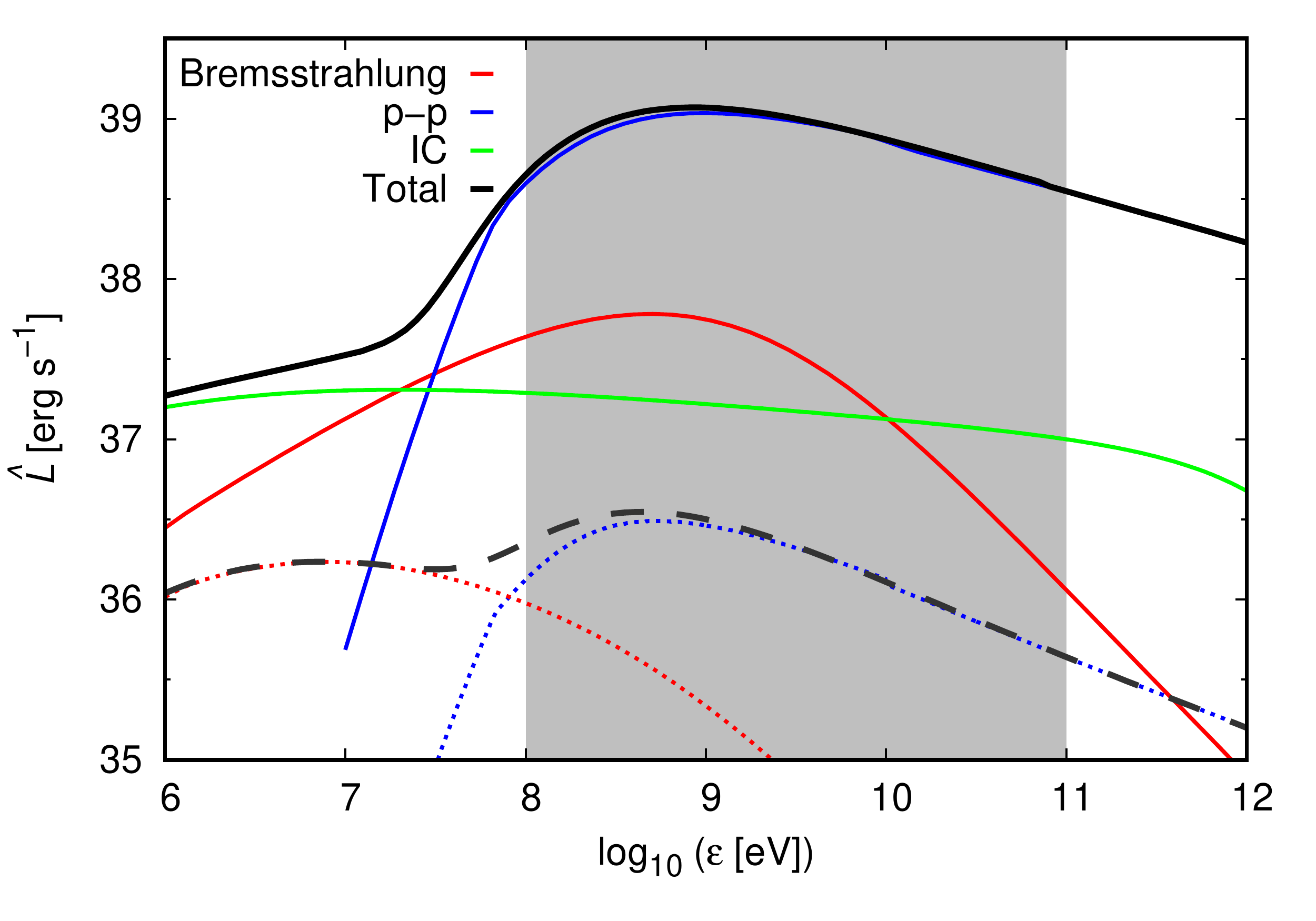}
  \caption{SEDs per unit SFR for galaxies in scenarios B0 (left panel) and K0 (right panel). We plot the luminosity divided by the SFR, $\hat{L} = \varepsilon L_\varepsilon (\dot{M}_* / \mathrm{M}_\odot \, \mathrm{yr}^{-1})^{-1}$, with $\varepsilon$ the photon energy and $L_\varepsilon$ the specific luminosity, to simplify the comparison between galaxies of different SFRs. Solid and dashed lines correspond to $\dot{M}_* = 200$ and $0.005 \, \mathrm{M}_\odot \, \mathrm{yr}^{-1}$, respectively. Colour lines are the individual contributions from different mechanisms (the colour code is the same as in Figs.~\ref{protoncooling} and \ref{electroncooling}), whereas black lines show the total SED. The grey shaded region is the \textit{Fermi} energy range. Cooling through $p$-$p$ dominates always in this range, and shows a strong supra linear behaviour with SFR.}
  \label{seds}
\end{figure*}

 \begin{figure}[ht]
  \centering
  \includegraphics[width=0.49 \textwidth]{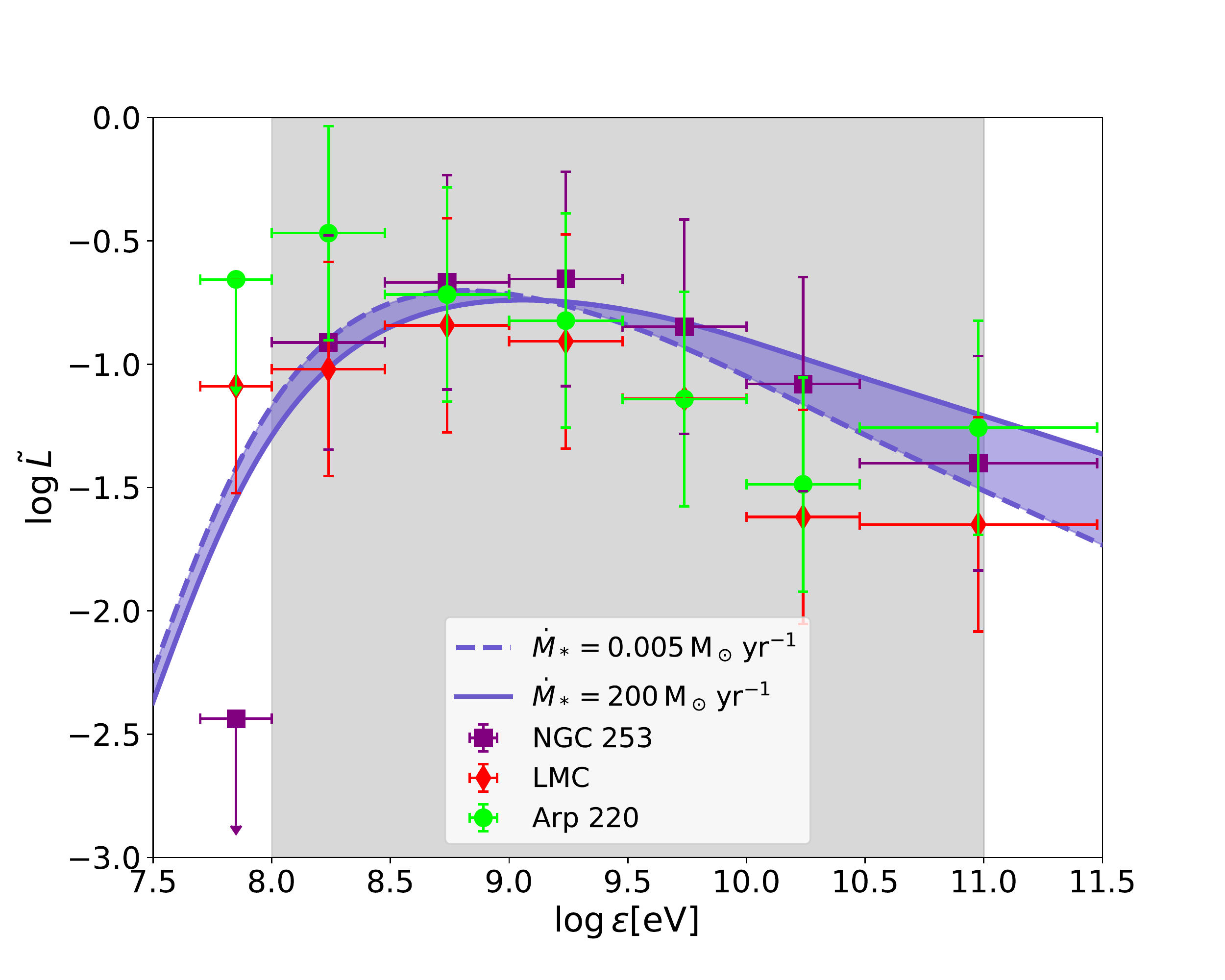}
  \caption{Observed normalised SEDs ($\tilde{L} = \varepsilon L_\varepsilon \, L_\gamma^{-1}$) of three galaxies spanning the whole SFR range. Model galaxies with $\dot{M}_* = 0.005\,\mathrm{M}_\odot\,\mathrm{yr}^{-1}$ (blue dashed line) and $\dot{M}_* = 200\,\mathrm{M}_\odot\,\mathrm{yr}^{-1}$ (blue solid line) in our scenario K0 are also plotted. The blue shaded region is the interpolation of the model to SFRs between these values. Our modelled SEDs agree fairly well with observed data within the \textit{Fermi} energy range (grey shaded region).}
  \label{sedsdata}
\end{figure}

The SED for the two extreme SFRs can be seen in Fig.~\ref{seds} for base scenarios K0 and B0. To compare the emission of galaxies with different SFRs, we normalise the SED dividing the luminosity by the SFR. Within the \textit{Fermi} energy range (the grey shaded region in Fig.~\ref{seds}), the SED comprises contributions IC, Bremsstrahlung and $p$-$p$. Although IC emission grows with SFR faster than $p$-$p$, the latter dominates over the entire SFR range. This result agrees with those of previous authors \citep[e.g.][]{2013ApJ...762...29L,2019MNRAS.487..168P}.

The prevalence of $p$-$p$ radiation is due to two facts. First, the hadron-to-lepton ratio for CRs in SFGs is high (we adopted $a = 50$, see Sect.~\ref{model}). Second, according to our model, most of the power of CR electrons is not radiated but rather lost by ionization. The maximum energy they radiate is $\sim 26\%$ of the total energy injected in them. The case of protons is different, since ionization losses are negligible, and only escape processes compete with radiation losses.

At high SFRs, the $\gamma$-ray spectrum is similar in both base scenarios (K0 and B0). The spectral index for K0 is $\alpha^*\sim -2.3$, a little bit steeper than that for B0 ($\alpha^*\sim -2.2$), because of the strong diffusion. These values agree with the observed spectral indices of SBGs \citep{ajello2020}. As SFR decreases, proton leakage begins to gain relevance. In the scenario B0 particles escape by advection, keeping $\alpha^*$ unchanged. In K0, protons escape by diffusion, steepening $\alpha^*$ to $\sim -2.4$ at $\dot{M}_* = 0.005\, \mathrm{M}_{\odot} \, \mathrm{yr}^{-1}$. Fig~\ref{sedsdata} shows that the agreement between modelled and observed SEDs is fairly good.

\subsection{The $L_\gamma$--SFR relation}

\begin{figure*}[t]
  \centering
  \includegraphics[angle=0, width=0.49 \textwidth]{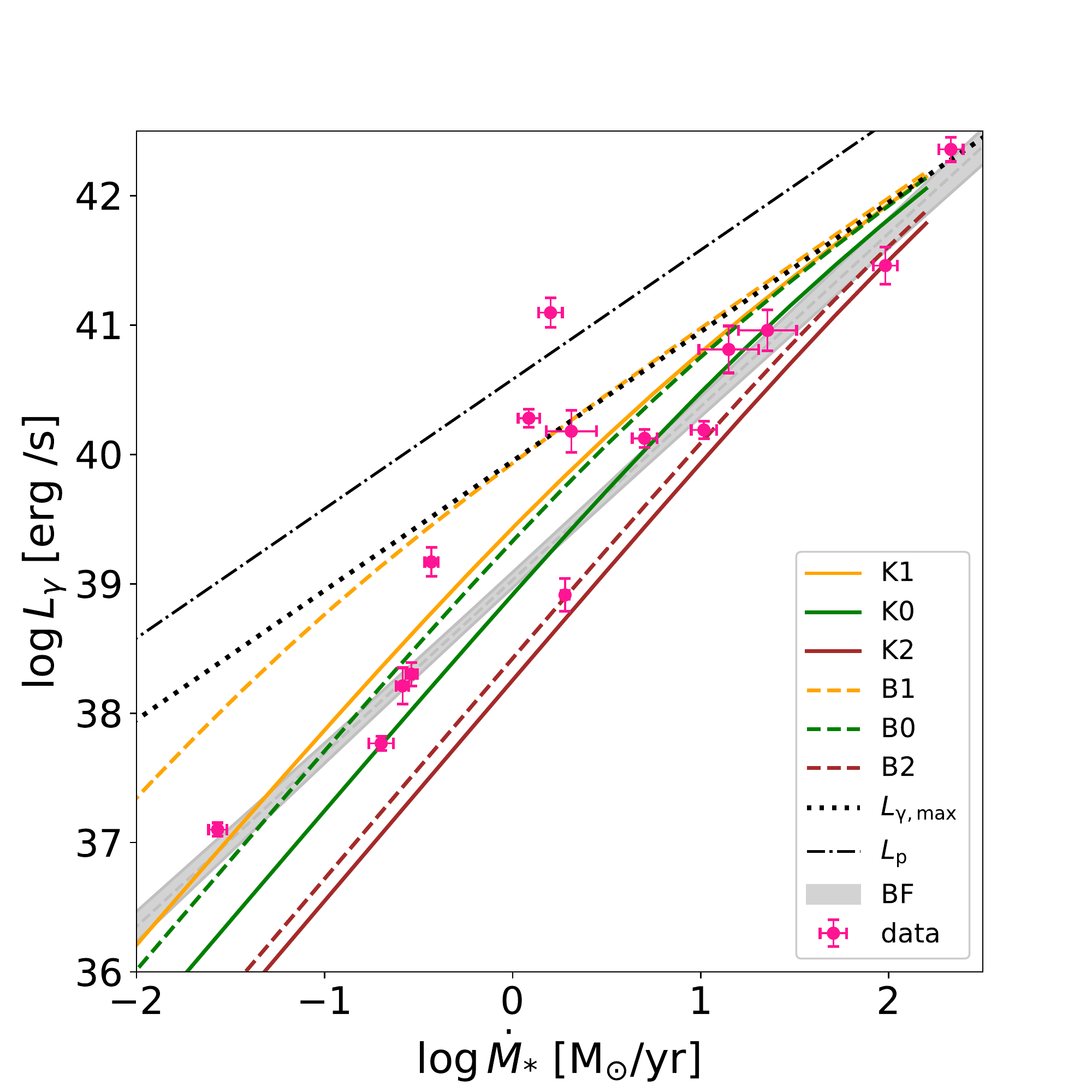}
  \includegraphics[angle=0, width=0.49 \textwidth]{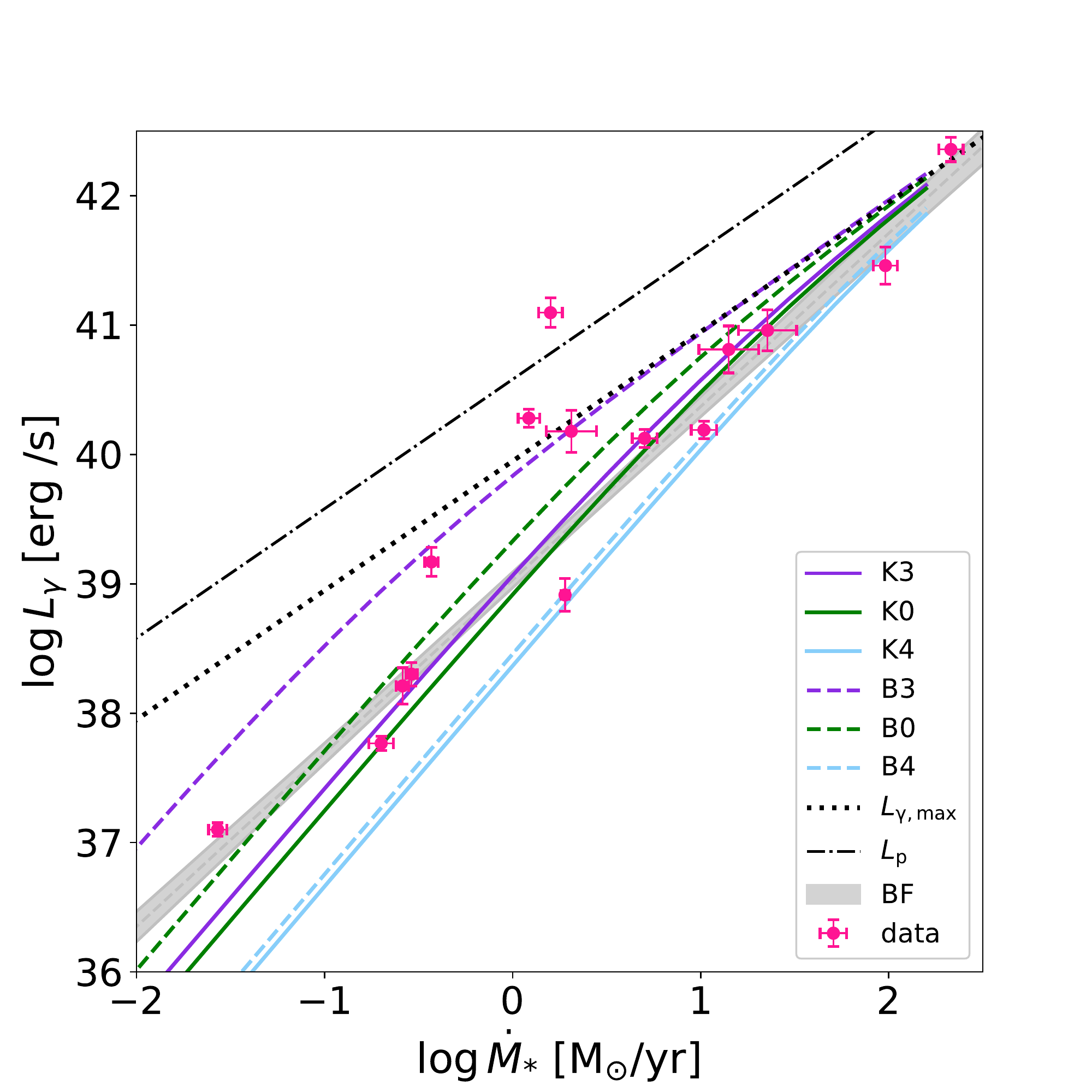} 
    \caption{The $L_{\gamma}$--SFR relation for our scenarios 0, 1, and 2 ($R = 1,\,0.1,\,5\,\mathrm{kpc}$, left panel), and 0, 3, and 4 ($v_\mathrm{w} = 400,\,40,\,4000\,\mathrm{km\,s}^{-1}$, right panel). Scenarios with both Kolmogorov (solid lines) and Bohm (dashed lines) diffusion prescriptions are shown. The black dotted line is the genuine calorimetric limit given by Eq.~\ref{calolimit} (see text), whereas the grey dot-dashed line represents all the luminosity available in relativistic protons for each SFR. The grey shaded band is the $1\sigma$ confidence region of the fit to the BF sample (See Sec.~\ref{data}). Data are consistent with scenarios that assume small-sized CR cooling regions and mild to high velocities, but only at high SFRs. At low SFRs, all models fail to describe the observed trend.}
  \label{modelrelation}
\end{figure*}

In Fig.~\ref{modelrelation} (left panel) we show the model $L_\gamma$--SFR relation for scenarios with different values of $R$, the radius of the region where CRs cool. $L_\gamma$ is computed by integrating the SEDs between 0.1 and $100\,\mathrm{GeV}$, for the twenty model galaxies of each scenario. We compare the model relations with that derived from the BF sample (see Sect.~\ref{data}).

We have shown in Sec.~\ref{resultsseds} that in the \textit{Fermi} energy range, the main contribution to $L_\gamma$ in our model is due to $p$-$p$ radiation. Therefore, the $L_\gamma$--SFR relation is regulated by the leakage of protons from the CR cooling region. The absolute maximum power available for $\gamma$-ray radiation is the CR luminosity of Eq.~\ref{eq:Lcr}, which scales linearly with the SFR (grey dot-dashed lines in Fig.~\ref{modelrelation}). However, this limit is unreachable, as only $\sim 33\%$ of the proton energy can be transformed into $\gamma$-ray photons \citep{Kelner2006} and, according to the SEDs produced by our models, only $\sim 76\%$ of the $\gamma$-ray luminosity is emitted in the \textit{Fermi} band. This gives a more genuine limit for the $\gamma$-ray luminosity of model galaxies (black dotted lines in Fig.~\ref{modelrelation}), 

\begin{equation}
    L_{\gamma,\mathrm{max}} [\mathrm{erg\,s}^{-1}] = 8.38\,\times 10^{39} \dot{M}_* [\mathrm{M}_\odot\,\mathrm{yr}^{-1}].
    \label{calolimit}
\end{equation}

\noindent
The departure of the emission from this limit,  $\rho = L_\gamma / L_{\gamma,\mathrm{max}}$, is therefore a measure of the ratio between radiated and available power, or calorimetric ratio. From the results of Sec.~\ref{data}, for the observed data $\rho_\mathrm{obs} \propto \dot{M}_*^{0.34 \pm 0.05}$. It is noteworthy that all galaxies, except \object{NGC 3424} and \object{NGC 4945} (which are outliers in other senses, see Sec.~\ref{data}), lie below $L_{\gamma,\mathrm{max}}$ to within observational uncertainties. This result is not expected a priori, since the computation of $L_{\gamma,\mathrm{max}}$ includes model-dependent factors. Therefore, it suggests that the model effectively captures the relevant physics of the problem.

Fig.~\ref{modelrelation} (left panel) shows that for all values of $R$, model galaxies approach the genuine \textit{Fermi} limit at high SFRs ($\rho \to 1$), whereas at lower SFRs they depart from it. The separation increases in a monotonic way as the SFR decreases. This is a density effect; high-SFR systems have higher ISM densities by construction, making $p$-$p$ cooling more efficient and dominant over escape mechanisms. On the other hand, at low SFRs, the density is not high enough to prevent the escape of an important fraction of the proton population, leaving less energy to be transformed into $\gamma$-rays ($\rho \to 0$). This result is consistent with previous works \citep{2013ApJ...762...29L,2018MNRAS.474.4073W,2019MNRAS.487..168P}.

At high SFRs ($\log \dot{M}_*\,[\mathrm{M}_\odot\,\mathrm{yr}^{-1}] \gtrsim 0.5$), the observed trend is well reproduced by our base scenario K0 ($R = 1\, \mathrm{kpc}$, Kolmogorov diffusion). The Bohm diffusion recipe (B0) shows a similar trend, but displaced from the locus of the observed galaxies. We recall that in B0 proton escape is driven by advection, at a slower rate. Therefore, B0 results in a higher calorimetric ratio than that of K0. Scenarios with larger radii (K2, B2, $R = 5\,\mathrm{kpc}$) underpredict the calorimetric ratio, because although larger systems have lower escape rates ($\tau_\mathrm{adv} \propto R$, $\tau_\mathrm{diff}\propto R^{2}$), they also have lower densities (at fixed SFR, $\tau_\mathrm{pp} \propto n \propto R^{-2.42}$) that make $p$-$p$ cooling much less efficient. These scenarios also show a steeper increase of $\rho$ with SFR than the one observed. On the other hand, the opposite is true for scenarios with small radii (K1, B1, $R = 0.1\,\mathrm{kpc}$). These are efficient calorimeters, and present $\rho(\dot{M}_*)$ trends shallower than that observed in our sample. Particularly, for Bohm diffusion (B1) we obtain $\rho \approx 1$ in the whole high-SFR range, which means $L_\gamma \propto \dot{M}_*$. To summarise, at high SFRs our model is consistent with kiloparsec-sized CR cooling regions. Smaller regions are not ruled out, but require that diffusion proceeds in the Kolmogorov regime.

At low SFRs ($\log \dot{M}_*\,[\mathrm{M}_\odot\,\mathrm{yr}^{-1}] \lesssim -0,5$), all scenarios fail to reproduce the observed trend. In all cases the $L_\gamma$--SFR relation is steeper than that observed. This suggests that the model overestimates the relative strength of escape with respect to $p$-$p$ losses, for any size of the system.

To assess the effects of variations in the galactic wind velocity, in Fig.~\ref{modelrelation} (right panel) we show the same relations of left panel, but for scenarios 3 and 4. As expected, higher wind velocities increase the escape, lowering and steepening the curves with respect to scenarios K0 and B0. On the contrary, lower velocities move the curves closer to the calorimetric limit, and make them shallower. At high SFRs, our model is consistent with mild to high wind velocities of several hundreds of kilometres per second. Slower winds cannot be discarded, but require the presence of Kolmogorov diffusion. Once again, at low SFRs, all scenarios fail to reproduce the observed trend, the curves being steeper than the relation determined from observations.

Finally, variations in the model magnetic field (scenarios 5 and 6) change the diffusion rate in the Bohm case, and synchrotron losses. We have already proved both processes to produce negligible effects. The results of scenarios 5 and 6 (for both diffusion prescriptions) show that the $L_{\gamma}$--SFR relation is not affected by changing the magnetic field, therefore, we do not discuss the latter further.

It is noteworthy that all scenarios fail in the same way at low SFRs. The $L_\gamma$--SFR relation shows a remarkably constant power-law index $m = 1.71$ in this SFR range, for all scenarios. This is different from the value $m = 1.35 \pm 0.05$ obtained from observations. We defer a thorough discussion of this result to Sec.~\ref{conclusions}, noting that only four galaxies lie in this SFR range, and that three of them (all but the \object{SMC}) seem to follow the model trend more closely.

\section{Discussion and conclusions}
\label{conclusions}

In this work we have compiled from the literature a near-homogeneous sample of distances and observed fluxes in the FUV, H$\alpha$, and IR, for all SFGs detected in $\gamma$ rays by \textit{Fermi}. We have used these data to obtain a self-consistent set of SFRs, and $\gamma$-ray and IR luminosities, to probe possible biases present in the $L_\gamma$--SFR correlation found by previous authors \citep{2012ApJ...755..164A,ajello2020}. Our work improves on that of \citet{ajello2020} by including a CR emission model describing this correlation for their full data set. Previous works use smaller data sets, although they develop more refined models \citep[e.g.,][]{2017ApJ...847L..13P} or explore different aspects of the correlation, such as the radio emission of SFGs \citep[e.g.,][]{2012ApJ...755..164A}.

Using the constructed sample, we have shown that $L_\mathrm{IR}$ consistently underestimates the SFR of $\gamma$-ray emitting galaxies for $\dot{M}_\odot \lesssim 1\,\mathrm{M}_\odot\,\mathrm{yr}^{-1}$. Although this is a known result for the general galaxy population \citep{2008ApJS..178..247K,2012ARA&A..50..531K}, we stress it because it produces a bias in the $L_\gamma$--$L_\mathrm{IR}$ correlation, usually not corrected for \citep[but see][]{2017ApJ...847L..13P,2019ApJ...874..173Z}. We have quantified this bias, finding that the power-law index of the  $L_\gamma$--SFR relation is underestimated by 0.14 when $L_\mathrm{IR}$ is used as an SFR tracer. According to our data, the most probable value for this index is $1.35 \pm 0.05$, although the present sample is small (fifteen galaxies, with some outliers) and errors are relatively large, mainly due to large distance uncertainties. In any case, a linear $L_\gamma$--SFR relation is discarded with high confidence.

As a by-product, we have found that at intermediate SFRs $0.3 \lesssim \dot{M}_* \lesssim 3\,\mathrm{M}_\odot \, \mathrm{yr}^{-1}$, the dispersion of the $L_\gamma$--SFR relation is tens of times higher than outside this range. The MW is the only intermediate-SFR galaxy standing below the best-fit correlation, whereas the other four galaxies (\object{NGC 2403}, \object{NGC 3424}, \object{NGC 4945}, and \object{Circinus galaxy}) lie well above it. The MW and \object{NGC 4945} have almost the same SFR, but their luminosities differ by 1.37~dex. It might be objected that the MW $\gamma$-ray luminosity is modelled instead of actually measured, but the MW model fits $\gamma$-ray, radio, and CR data, therefore it is reliable, and the discrepancy cannot be attributed to this fact. On the other hand, all four objects lying above the fit have suspected or confirmed AGNs \citep{Yang2009,Gavazzi2011,Yaqoob2012,Peng2019}, whereas the MW has a starving black hole in its centre \citep{2002Natur.419..694S}. Therefore, we suggest that the large dispersion of the $L_\gamma$--SFR relation in the intermediate SFR range is due to an AGN component in $L_\gamma$, whose strength varies from galaxy to galaxy. We argue that this component is not observable in other SFR ranges because low-SFR spirals have no AGN, whereas in ULIRGs the AGN emission may be highly absorbed by the dense interstellar medium. At densities of $3\times 10^4\,\mathrm{cm}^{-3}$, typical of the innermost regions of SBG nuclei, pair production in the nuclear electric field of \ion{H}{i} would provide the required absorption mechanism for GeV photons \citep{2018PhRvD..98c0001T}. The confirmation of this claim, namely that AGNs contribute to the $\gamma$-ray luminosity of  intermediate-SFR galaxies, would require the development of a method to separate the stellar-population component from that of the AGN in $\gamma$-ray observations. Such a method would certainly improve the observed $L_\gamma$--SFR relation. 

A different possibility would be that $L_\gamma$ comprises in some galaxies contributions from other sources, such as halo super-bubbles created by strong winds like those modelled by \citet{2018A&A...616A..57R}. Alternative explanations would be that the power-law nature of the $L_\gamma$--SFR relation breaks at intermediate SFRs, producing a bump, or that normal spirals and SBGs follow different relations, which coexist in the mid-SFR range. In this case, extra observables would be needed to separate galaxies in both regimes. An exploration of this hypothesis will be addressed in a future work.

To explore the physics behind the observed $L_\gamma$--SFR relation, we developed a simple leaky-box model to compute the CR populations in SFGs, and their high-energy emission. Our \textit{Fermi}-band model SEDs agree with those observed \citep{ajello2020}. The emission is dominated by $p$-$p$ inelastic scattering, in agreement with previous works \citep{2013ApJ...762...29L,2019MNRAS.487..168P}, the leptonic radiation being negligible. The genuine calorimetric limit resulting from our model (which takes into account that only a fraction of the proton energy can be radiated) closely matches the emission of the highest-SFR systems, indicating that the model includes the relevant physics of $\gamma$-ray emission in SFGs.

Our model describes fairly well the $L_\gamma$--SFR relation at high SFRs, provided that CR cooling occurs in kiloparsec-sized regions, and galactic winds blow at velocities of several hundreds of kilometres per second. This is in line with the findings of \citet{2019ApJ...872...16D}, that star formation regions in galaxies are smaller than galactic discs, by a factor of almost two. In this regard, our model disagrees with those previously used to compute the $\gamma$-ray emission of SBGs, which assume that the emission arises from spherical regions of $\sim 0.2\,\mathrm{kpc}$ in radius \citep[e.g.,][]{2019MNRAS.487..168P}. The discrepancy may be due to the different geometries used, as different sizes are required to reach a similar proton density, which is the variable that controls $p$-$p$ emission. Our model disagrees also with the simulations of \citet{2017ApJ...847L..13P}, who show in their Fig.~1 $\gamma$-ray emission extending to $\sim 10\,\mathrm{kpc}$-sized regions in galactic discs. The solution of these discrepancies requires spatially resolved $\gamma$-ray observations, not available at present for galaxies beyond the Local Group.

At low SFRs, our model predicts a steeper trend than observed, implying that the increase of particle escape with SFR has been overestimated. The power-law index of our model relation as $\dot{M}_* \to 0$ reaches a limit of $1.71$, because for small densities (implied by the K-S law), $L_\gamma \propto L_\mathrm{CR} \tau_\mathrm{esc} \tau_\mathrm{pp}^{-1} \propto \dot{M}_*^{1+0+0.71}$, from Eqs.~\ref{eq:Lcr}, \ref{taupp}, and \ref{eqdensity} (once again, the K-S law), and the fact that escape does not depend on SFR due to the constancy of $R$ and $v_\mathrm{w}$. It is noteworthy that, assuming that advection dominates through neutral winds like those observed by \citet{Martin2005} and \citet{Weiner2009}, we would obtain $\tau_\mathrm{esc} \propto \dot{M}_*^{-0.35}$, rendering $L_\gamma \propto \dot{M}_*^{1.36}$, in excellent agreement the value of $1.35 \pm 0.05$ derived by us from observations. In our model, advection dominates in scenarios with Bohm diffusion, or in fast-wind scenarios with any diffusion prescription. However, as explained in Sec.~\ref{model}, our Kolmogorov recipe represents the fastest diffusion, which is a very extreme situation. A lower normalization of the Kolmogorov diffusion coefficient, such as that used by \citet{2019MNRAS.487..168P} in their model A, would provide and advection dominated regime also in our scenarios with Kolmogorov diffusion.
This is the most plausible explanation of the observed trend at low SFRs: the $L_\gamma$--SFR relation may be driven by the combination of CR luminosity, the K-S law, and an advection-dominated escape regime with an SFR-dependent wind velocity. This explanation would also alleviate the tension of our model with present limits for the MW wind velocity, restricted to some tens of kilometres per second \citep[e.g.,][]{2007ARNPS..57..285S}. SFR-dependent wind velocities can arise  in different scenarios \citep[e.g.,][and references therein]{Veilleux2005}, including CR-driven winds \citep[e.g.,][]{2016ApJ...816L..19G}; these scenarios are a promising clue to explore in follow-up works.

To summarise, we have provided strong evidence for the existence of a bias in previous determinations of the $L_\gamma$--SFR relation of SFGs, due to the use of IR luminosity as an SFR tracer. A quantitative estimation of the actual power-law index of this relation is $1.35 \pm 0.05$. Physically-motivated, population-oriented models of $\gamma$-ray emission show that the unbiased relation can be explained at high SFRs by assuming that the CR cooling region is kiloparsec-sized, and pervaded by mild to fast winds. Combined with previous results about the scaling of wind velocity with SFR, our work provides support to advection as the dominant CR escape mechanism in low-SFR galaxies. The question of whether the emission of normal SFGs and SBGs is based on the same relevant physics, or if two different relations apply, is still open. A step forward in the comprehension of $\gamma$-ray emission by the stellar populations of SFGs requires further observations to enlarge the present sample and reduce measurement errors, and a reliable technique to disentangle the stellar contribution from that of the AGN, if present. On the theoretical side, population-oriented models may provide further insight into the conditions prevailing in SFGs, as far as the accuracy of present SFR scaling relations is improved, and new ones are unveiled.

\begin{acknowledgements}
    PK and LJP acknowledge support from Argentine CONICET (PIP-2014-00265).
      Part of this work was supported by the German
      \emph{Deut\-sche For\-schungs\-ge\-mein\-schaft, DFG\/} project
      number Ts~17/2--1. JFAC acknowledges support from UNRN funds (project 40-C-691). JFAC is a staff researcher of the CONICET. GER was supported by the Argentine agency CONICET (PIP 2014-00338) and the Spanish Ministerio de Economía y Competitividad (MINECO/FEDER, UE) under grant AYA2016-76012-C3-1-P.
      This research has made use of the NASA/IPAC Extragalactic Database (NED), which is funded by the National Aeronautics and Space Administration and operated by the California Institute of Technology.
\end{acknowledgements}

%
   \bibliographystyle{aa} 
   \bibliography{ref.bib} 
%

%



\end{document}